\documentclass[fleqn,usenatbib]{mnras}

\usepackage{newtxtext,newtxmath}
\usepackage[T1]{fontenc}

\DeclareRobustCommand{\VAN}[3]{#2}
\let\VANthebibliography\thebibliography
\def\thebibliography{\DeclareRobustCommand{\VAN}[3]{##3}\VANthebibliography}

\usepackage{graphicx}
\usepackage{amsmath}
\usepackage{booktabs}
\usepackage{threeparttable}
\usepackage{multirow}

\title[Extremely eccentric BNSs in the Milky Way]{Formation and Detectability of PSR\,J0653+0443-like Extremely Eccentric Binary Neutron Stars in the Milky Way}

\author[Chen et al.]{
Zhiwei Chen,$^{1,2}$
Jihui Zhang,$^{1,3}$
Youjun Lu,$^{3,1}$\thanks{E-mail: luyj@nao.cas.cn}
\\
$^{1}$National Astronomical Observatories, Chinese Academy of Sciences, 20A Datun Road, Beijing 100101, China\\
$^{2}$Institut f\"ur Theoretische Physik, Goethe Universit\"at, Max-von-Laue-Str. 1, 60438 Frankfurt am Main, Germany\\
$^{3}$School of Astronomy and Space Sciences, University of Chinese Academy of Sciences, 19A Yuquan Road, Beijing 100049, China\\
}

\date{Accepted XXX. Received YYY; in original form ZZZ}
\pubyear{\the\year{}}

\begin{document}
\label{firstpage}
\pagerange{\pageref{firstpage}--\pageref{lastpage}}
\maketitle

\begin{abstract}
Recently, the Five-hundred-meter-Aperture-Spherical-radio-Telescope (FAST) Galactic Plane Pulsar Snapshot (GPPS) survey discovers an extraordinary binary neutron star (BNS) system, PSR\,J0653+0443, possessing the highest known orbital eccentricity ($e\approx0.992$) among Galactic BNSs \citep{yang_np}, {which provides a unique probe into the formation and evolution of BNSs.} In this Letter, we perform compact binary population synthesis calculations with self-consistent neutron star spin evolution to model extremely eccentric BNS population detectable by the FAST GPPS survey, explicitly incorporating observational selection effects. We find that BNSs with $e>0.99$ constitute only $\sim0.6\%$ of the detectable population, with an expected {detection number of $\lambda_{\mathrm{GPPS}}\approx0.31$} {over the full planned footprint}. Thus, although rare, systems as eccentric as PSR\,J0653+0443 are compatible with standard binary evolution. We further demonstrate that extremely eccentric BNSs arise preferentially from core-collapse supernovae, whereas imposing $P\lesssim0.1$\,s  select older, more frequently recycled systems, which naturally reproduce the observed properties of PSR\,J0653+0443. We further predict that the {completed GPPS sample has a $\sim15\%$} chance of finding another $e>0.99$ BNS, but a much higher $\sim88\%$ chance for $e>0.90$ systems. {By contrast, future facilities are expected to enlarge the detectable highly eccentric BNS population, with SKA potentially detecting PSR\,J0653+0443-like systems and sub-mHz gravitational-wave detectors probing a larger population with $e>0.9$, enabling much more in-depth population-level
constraints.}
\end{abstract}

\begin{keywords}
binaries: general -- stars: neutron -- pulsars: general -- stars: evolution -- gravitational waves
\end{keywords}

\section{Introduction}

Binary Neutron Stars (BNSs) are among the most important targets for both electromagnetic (EM) and gravitational-wave (GW) observations, of which physical and orbital properties can be measured accurately through pulsar timing. They encode rich information about the evolution of massive binary stars and the properties of neutron stars \citep[e.g.,][]{2018MNRAS.481.4009V, 2019MNRAS.486.3213A, 2025arXiv250814397C}, while also serving as natural laboratories for studying strong-field gravity and relativistic processes under extreme physical conditions \citep[e.g.,][]{1982ApJ...253..908T, 2016ApJ...829...55W, 2025arXiv251012506M}. The discovery of the Hulse-Taylor BNS system ( B 1913+16) \citep{1975ApJ...195L..51H} provided the first indirect evidence for GWs and revolutionized our understanding of compact binaries. Subsequent radio observation on pulsars have unveiled about $\sim 30$ BNS systems in our Milky way (MW) through precise pulsar timing techniques \citep[e.g.,][see also Supplementary Table B1.]{2024MNRAS.530.1506S,2025RAA....25a4003W}; These systems can serve as valuable probes for   aspects of astrophysics and fundamental physics \citep[e.g.,][]{2019MNRAS.482.3249Z,2024MNRAS.527.1101G}. 

The Five-hundred-meter Aperture Spherical radio Telescope (FAST; \citet{2011IJMPD..20..989N}), with its unprecedented 300-meter effective aperture, boasts the largest collecting area of any operational radio telescope. Equipped with a 19-beam L-band receiver featuring a system temperature of ~20 K, FAST currently is the most sensitive radio telescope for pulsar detection. Capitalizing on this capability, \citet{2021RAA....21..107H} initiated the Galactic Plane Pulsar Snapshot (GPPS) survey, targeting pulsar discoveries within the Galactic latitude of $\pm 10^{\circ}$ from the Galactic plane. Over its five-year operational span to date, the FAST GPPS survey has uncovered nearly a thousand new pulsars \citep[e.g.,][]{2023MNRAS.526.2645S, 2025RAA....25a4001H, 2025RAA....25a4002Y, 2025arXiv250614519J}, among which several are BNS systems \citep[e.g.,][]{2024MNRAS.530.1506S,2025RAA....25a4003W,2026arXiv260727333Y}.

Recently, FAST GPPS survey discovered an extraordinary BNS system, J0653+0443 (GPPS0539), distinguished by its remarkable eccentricity of $e\sim 0.992387$, setting a new record as the most eccentric compact binary system ever observed \citep{yang_np}. {The system has a spin period $P\simeq47.5$ ms,
$\dot{P}\simeq1.14\times10^{-19}$, and an orbital period
$P_{\rm orb}\simeq79.1$ days, with pulsar and companion masses of
$M_{\rm psr}\simeq1.408\,M_\odot$ and
$M_{\rm c}\simeq 1.399\,M_\odot$.}
Interestingly, while  previous radio observations have identified one BNS systems with high eccentricity, i.e., J0514–4002A \citep[$e\sim 0.888$,][]{Ridolfi+2019MNRAS.490.3860R}, its presence in globular clusters allows the high eccentricity to be naturally explained by dynamical interactions in dense stellar environments \citep[e.g.,][]{2019MNRAS.488...47F,2020ApJ...898..152S}. In contrast, J0653+0443 is located in the MW field $\rm (gl,gb)=(208.9^{\circ},2.51^{\circ})$  rather than in globular clusters, indicating that it is more likely formed via the isolated evolution of massive binary stars \citep[ e.g.,  ][]{2002MNRAS.329..897H,2016ApJ...819..108B, 2022LRR....25....1M}. Therefore, observations of this extremely eccentric BNS system may place stringent constraints on the formation and evolution of BNSs.

In this letter, we systematically investigate the population of extremely eccentric BNS systems as PSRJ0653+0443 (i.e., $e\gtrsim0.9$) detectable by GPPS survey, utilizing compact binary population synthesis (CBPS) with NS spin evolution. This letter is organized as follows. In section~\ref{sec:2}, we constrain our CBPS model with latest multi-messenger constraint on BNS systems and find the most compatible parameter settings, considering the selection effect. In section~\ref{sec:3}, we conduct a dedicated analysis on the properties of extremely eccentric BNS systems detectable by FAST GPPS survey. In Section~\ref{sec:4}, we present a  prospect on the future detectability of extremely eccentric BNS with FAST GPPS survey and square kilometer array (SKA). Conclusions and discussions are shown in Section~\ref{sec:5}.
     
\section{Model Constraint using FAST GPPS BNS}
\label{sec:2}

In this section, we first put constraints on our compact binary population synthesis (CBPS) model by using two latest observational results, i.e., (1) the local BNS merger rate density inferred from the LVK O4 observing run \citep{2026arXiv260527226T}, and (2) the orbital properties of Galactic BNS systems detectable by the FAST GPPS survey, including PSR J0653+0443. These observational constraints provide a direct diagnostic of the key parameters in the formation and evolution of BNS systems. 

We utilize an updated version of CBPS code \texttt{BSE} to simulate the formation and evolution of BNS systems, which is originally developed by \citet{2000MNRAS.315..543H,2002MNRAS.329..897H}, and later modified and extended by \citet{2010A&A...521A..85Y}, \citet{2022MNRAS.509.1557C,2025ApJ...980..181C}, and \citet{2025arXiv250814397C}. The initial condition for BNS simulation are set as the same in \citet{2025ApJ...980..181C}. During the subsequent binary evolution, the common envelope (CE) phase plays a central role in tightening the orbit and enabling the formation of compact binaries. In this work, we model the CE evolution using the widely adopted $\alpha$-formalism \citep{1984ApJ...277..355W}. In addition, to simultaneously explain the low BNS merger rate density and the existence of the short-orbital-period Galactic BNS systems, we follow our previous work \citet{2025arXiv250814397C} to adopt a ``median CE'' description by introducing $f_{\rm HG}$, the fraction of Hertzsprung-gap (HG) donors that are forced to merge in the CE stage. Note that here $f_{\rm HG}$ is only a simplified phenomenological prescription, while the HG-donor survival during CE evolution should depend on various specific physical settings. Thus, the fixed $f_{\rm HG}$ adopted here should only be interpreted as an effective interpolation between the optimistic and pessimistic CE limits in the literature \citep[e.g.,][]{2007ApJ...662..504B, 2012ApJ...759...52D, 2013ApJ...779...72D, 2018MNRAS.480.2011G, 2020ApJ...898..152S, 2024ApJ...976...24B} rather than a unique physical CE model.

When massive stars undergo supernova explosions (SNe), newly formed NSs receive natal kicks, which are commonly modeled using a Maxwellian distribution inferred from proper-motion measurements of young isolated pulsars \citep{2005MNRAS.360..974H}. Notably, the value of dispersion $\sigma_{\rm k}$ of the Maxwellian distribution is dependent on the SNe mechanism \citep[e.g.,][]{2018MNRAS.481.4009V,2022MNRAS.509.1557C}. Specifically, the kick dispersion for electron-capture SNe \citep[ECSNe,][]{1980PASJ...32..303M} and ultra-stripped SNe \citep[USSNe,][]{2015MNRAS.451.2123T} might be smaller than that for core-collapse SNe \citep[CCSNe,][]{2002MNRAS.329..897H,2005MNRAS.360..974H}. After the formation of the BNS system, we follow \citet{2025ApJ...980..181C} to simulate the spin evolution of NS by considering several different phases in high-mass X-ray binary stage, in order to determine their radio lifetimes and observational visibility. A detailed description of these prescriptions is provided in
Supplementary Appendix A.1 of the online Supplementary Material.

To connect the synthetic population to observations, we incorporate the detection selection effect of the FAST GPPS survey into the population synthesis framework \citep[e.g.,][]{2020MNRAS.494.1587C, 2023MNRAS.526.2210S}. We account for both the sky coverage of the survey and the radio detectability of pulsars \citep[e.g.,][]{2011IJMPD..20..989N, 2021RAA....21..107H, 2024MNRAS.530.1506S, 2025RAA....25a4003W}, including the requirement that at least one NS remains radio-loud for FAST GPPS survey. This step allows the simulated population to be compared directly with the observed FAST BNS sample under realistic detection conditions. The implementation of the selection procedure is described in Appendix A.2 of the online Supplementary Material.

Using the above prescriptions, we generate synthetic BNS populations following both cosmic and Galactic star formation histories as in \citet{2022MNRAS.509.1557C}. We then perform the model constraint by exploring a grid of CBPS models spanning different values of  $\alpha$ and  $f_{\rm HG}$, which may strongly influence both the formation rate and orbital properties of BNS systems (see Supplementary Table A2). For each model ${M}_i$, we compute the local BNS merger rate density $R^{i}_0$ and evaluate the likelihood ${\rm log} L_i$ of reproducing the observed orbital period–eccentricity $P_{\rm orb}-e$ distribution of Galactic BNS systems detectable by FAST \citep[e.g.,][]{2018MNRAS.481.4009V,2025arXiv250814397C}. The details on the calculation is shown in  Appendix A.3 of the online Supplementary Material. 

\begin{figure}
\centering
\includegraphics[width=\columnwidth]{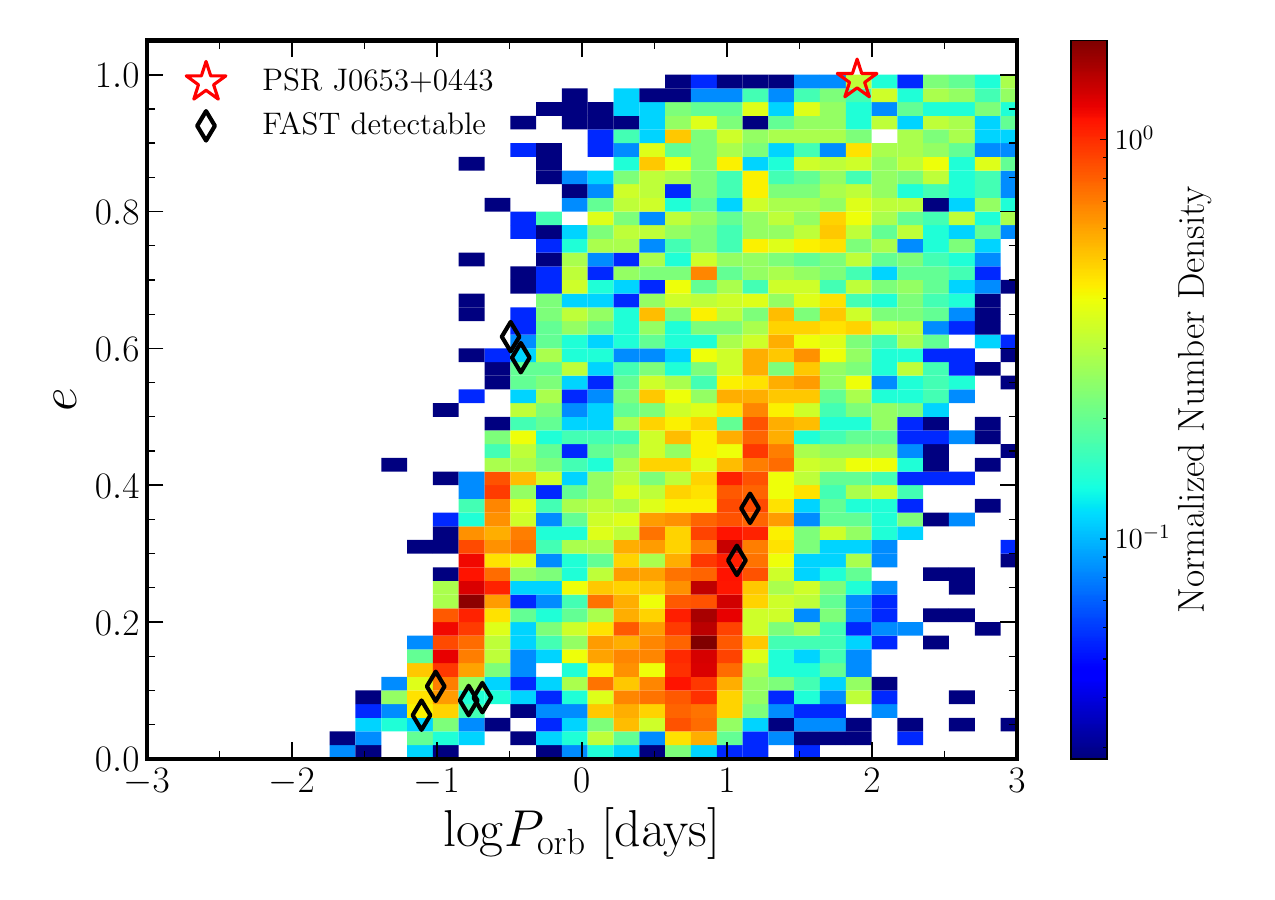}
\caption{The orbital period-eccentricity $P_{\rm orb}-e$ distribution of synthetic Galactic BNS with the most compatible model $M_1$ ($\alpha = 5$ and $f_{\rm HG} = 0.8$ ), detectable for FAST GPPS survey. For mock BNSs in this figure, both the localisation and radio selection effect have been taken into account. The red open star represent the newly found high eccentric pulsar PSRJ0653+0443, and the rest black open diamonds represent other {eight} field BNSs detectable with FAST GPPS in ATNF Catalog (see Supplementary Table B1). 
}
\noindent{Alt text:}
Two-dimensional distribution of orbital period and eccentricity for
synthetic Galactic BNSs detectable by the FAST GPPS survey. The
observed field BNSs are overplotted, with PSR J0653+0443 located at
the extreme high-eccentricity tail of the distribution.

\label{fig:f1}
\end{figure}

Among the models considered in this work (see Supplementary Table A2), we find that $M_1$ with $\alpha = 5$ and $f_{\rm HG} = 0.8$ still provides the best agreement with both observational constraints, consistent with our previous results in \citet{2025arXiv250814397C}. The predicted merger rate density $R_0\sim 149~\rm Gpc^{-3}yr^{-1}$ is consistent with current observational estimates. Moreover, as shown in Figure~\ref{fig:f1}, the simulated $P_{\rm orb}-e$  distribution nicely reproduces the observed FAST BNS population with {the highest ${\rm log} L_i\sim -18.22$.} 
Hereafter we adopt this best model as the fiducial model in the following analysis. 

\section{Statistical Properties of Extremely Eccentric BNS}
\label{sec:3}

The discovery of PSR~J0653+0443, with a remarkably high orbital eccentricity of $e\sim 0.992387$, which may provide important insights of the formation and evolution of BNSs. While the calibrated CBPS model $M_1$ presented in Section~2 successfully reproduces the overall orbital properties of Galactic BNS systems detectable by the FAST GPPS survey and local merger rate density measured by LVK GW detection, systems with high eccentricity like PSR~J0653+0443 are predicted to be rare (see Figure~\ref{fig:f1}). In this section, we perform a dedicated statistical analysis on the properties of extremely eccentric BNS systems within the calibrated population model.  

\begin{table}
\caption{{Predicted detectability of eccentric BNSs in the FAST GPPS survey.} The first, second, and third rows show the detectable fraction $f_{\rm GPPS}$, the expected number of detections $\lambda_{\rm GPPS}$, and the probability of detecting at least one extremely eccentric BNS in the GPPS survey, respectively. The mock BNS samples are generated by the calibrated model $M_1$.}
\label{table:1}
\centering
\renewcommand{\arraystretch}{1.15}
\setlength{\tabcolsep}{8pt}
\begin{tabular}{lccc}
\hline\hline
 & $e>0.90$ & $e>0.95$ & $e>0.99$ \\
\hline
$f_{\rm GPPS}$                & 6.91\% & 3.32\% & 0.64\% \\
$\lambda_{\rm GPPS}$          & 3.32   & 1.59   & 0.31   \\
$P(N_{\rm GPPS}^{e}\geq 1)$   & 96.4\% & 79.6\% & 26.7\% \\
\hline
\end{tabular}
\end{table}

To quantify the rarity of extremely eccentric BNS systems, we compute the fraction of systems exceeding different eccentricity thresholds, i.e., $e>0.90, 0.95,$ and $0.99$, within the synthetic FAST GPPS-detectable population. As summarized in Table~\ref{table:1}, only a small fraction of detectable systems exhibit extreme eccentricities. In particular, systems with $e > 0.99$ constitute a tiny subset of the overall population, with a {fraction of $0.64\%$}, confirming that objects such as PSR J0653+0443 occupy the extreme tail of the Galactic BNS distribution. We further estimate the expected number of detectable systems above each eccentricity threshold. Specifically, we multiply the fraction by the total number of Galactic BNS systems expected to be detectable by the FAST GPPS survey {over the full planned footprint}, $N_{\rm GPPS} \sim 48$, as predicted by the fiducial $M_1$ model, yielding the expected number $\lambda_{\rm GPPS} = f_{\rm GPPS} \times N_{\rm GPPS}$ shown in the second row of Table~\ref{table:1}. {For the most extreme case ($e>0.99$), this corresponds to
$\lambda_{\rm GPPS}\simeq0.31$, indicating that fewer than one such
system is expected on average over the full planned GPPS footprint.} Given that the expected numbers are small, it is reasonable to model the detection counts as following Poisson statistics with mean $\lambda$. Under this assumption, the probability of detecting at least one extremely eccentric system is given by $P(N_{\rm GPPS}^{e} \geq 1) = 1 - \exp(-\lambda_{\rm GPPS})$, as listed in the third row of Table~\ref{table:1}. For $e>0.99$, {this probability is $P(N_{\rm GPPS}^{e}\ge1)\simeq 26.7\%$}. 
{Besides, we also verify that this conclusion is robust across the other high-likelihood models M2 and M3, which also predict that $e>0.99$ systems constitute only a small fraction of the GPPS-detectable BNS population (see Supplementary Appendix B).}
These results indicate that the discovery of an extreme system such as PSR J0653+0443 is statistically rare but not implausible within the standard binary evolution framework. Here we note that although BNS in dense star fields may experience complex dynamical interactions that enhance their eccentricity \citep[e.g.,][]{10.1093/mnras/173.3.729, 2019ApJ...880L...8A}, PSR J0653+0443 is located in the Galactic field, making the dynamical scenario unlikely.  

\begin{figure}
\centering
\includegraphics[width=1.0\columnwidth]{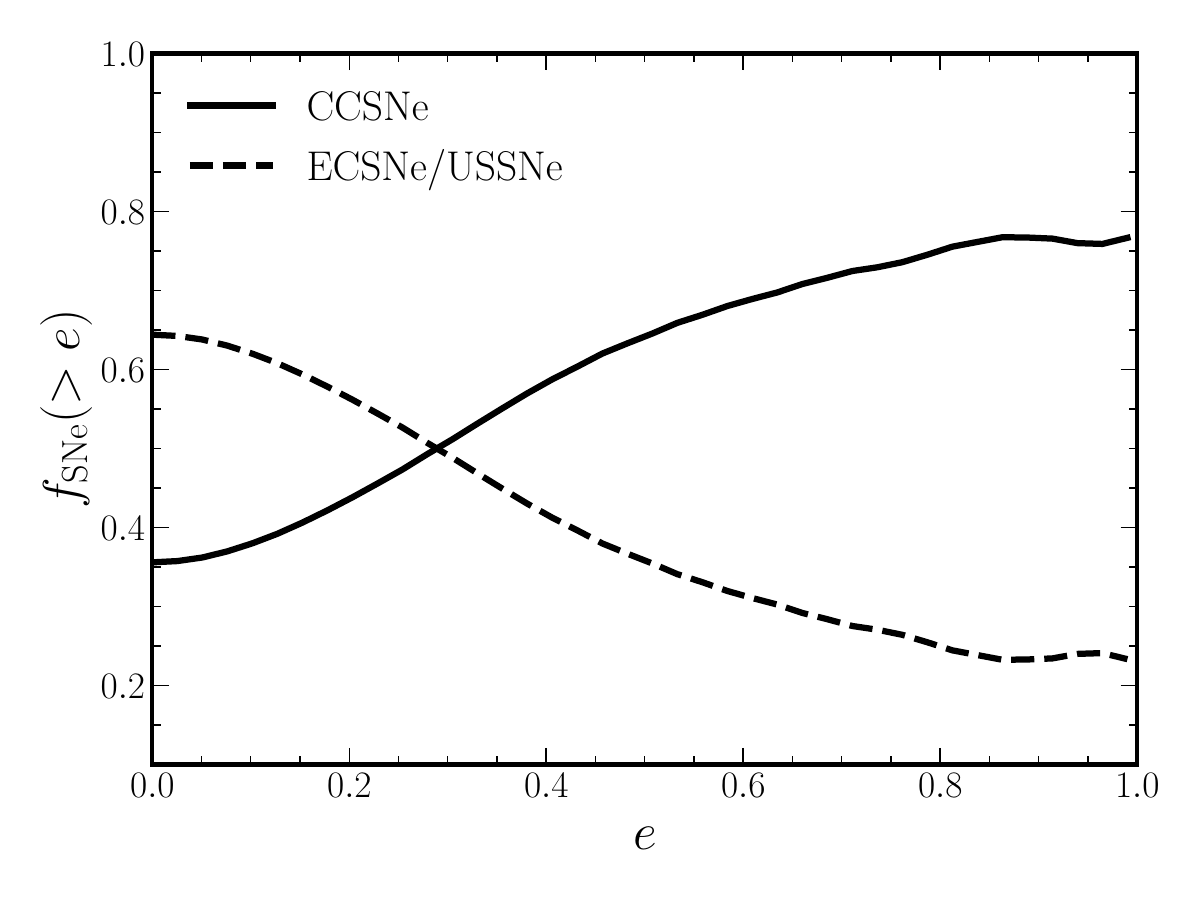}
\caption{Occupation fraction $f_{\rm SNe}(>e)$ of different SNe channels responsible for forming the second-born NS in BNS systems as a function of eccentricity threshold $e$. The solid and dashed line represent the results for CCSNe and ECSNe/USSNe respectively.
}
\noindent{Alt text:}
Line plot showing the occupation fractions of CCSN and ECSN/USSN formation channels as a function of eccentricity threshold. The CCSN fraction increases with increasing eccentricity, while the ECSN/USSN fraction decreases correspondingly.
\label{fig:f2}
\end{figure}

After quantifying the rarity of extremely eccentric BNS systems, we further investigate their physical properties. Figure~\ref{fig:f2} shows the occupation fraction $f_{\rm SNe}(>e)$ of different SNe channels responsible for forming the second-born NS in BNS systems as a function of eccentricity threshold $e$. Here we focus on the second SNe because it largely determines the post-explosion orbital configuration through the natal kick. A clear trend emerges that the fraction of CCSNe increases with eccentricity, while the contribution from ECSNe/USSNe decreases correspondingly. 
{For moderately eccentric systems ($e\lesssim0.30$)}, ECSNe/USSNe dominate the formation channel, whereas CCSNe become the primary channel for systems with $e\gtrsim0.9$. 
{
This transition indicates that extremely eccentric systems like J0653+0443 are preferentially associated with CCSNe channels, likely reflecting the combined effects of their typically larger natal kicks or more substantial SN mass loss.
} 
The distribution of SNe natal kick of extremely eccentric systems can be seen in  Appendix B of the online Supplementary Material.

A novel aspect of our CBPS analysis, compared with other work, is the self-consistent treatment of spin evolution by incorporating multiple detailed mechanisms during the formation processes of the binaries (see Supplementary Appendix A.1). We now investigate the spin evolution properties of extremely eccentric systems, which provide additional constraints on their evolutionary stage and observability.
{Here, we classify a pulsar as recycled if the system undergoes
Roche-lobe overflow after the CE phase. Wind accretion is
included in the spin evolution, but is not used as a criterion for
recycling.} {We find that only $\sim 14.0\%$ of the BNS systems} with $e>0.99$ have experienced stable Roche lobe overflow. However, when restricting the pulsar spin period to $P\lesssim0.1~\rm s$, as is the case for PSR J0653+0443, {the recycled fraction increases to $\sim 40\%$ for the same eccentricity range}. This naturally reconciles the recycled nature of PSR J0653+0443 with the statistical properties of the high-eccentricity population. The formation time of systems with $e>0.99$, defined as the elapsed time since the second supernova, {is about $\sim 163_{-156}^{+1168}\,{\rm Myr}$} (uncertainties represent the $16\%$th and $84\%$th percentiles). Applying the same spin-period cut {raises the median formation time to $\sim 1701_{-1378}^{+6452}\,{\rm Myr}$}, reflecting  the observational bias that short-period pulsars remain detectable over longer timescales, thereby allowing older systems to enter the observable sample. For reference, the characteristic age $\tau\sim P/2\dot{P}\sim 2400 \,{\rm Myr}$ inferred for PSR J0653+0443 is broadly compatible with our population predictions, {although $\tau$
is only a rough measurement of the time since the second supernova.} The corresponding distribution in the $P$–$\dot{ P}$ diagram and formation time distribution for extremely eccentric systems are shown in Supplementary Appendix B for reference. We caution here that the above estimates of the recycled fraction and formation time are model dependent. If future observations increase the local merger rate $R_0$, a smaller $\alpha$ may be favored, under which the population model would yield a larger recycled fraction and longer lifetimes for the extremely eccentric BNS systems.

In summary, our results reveal that extremely eccentric BNS systems tend to be younger and less frequently recycled than full detectable systems. However, once the additional short spin-period selection ($P\lesssim 0.1$\,s) is imposed, the observable high-eccentricity subsample shifts toward substantially older ages ($\sim 1\rm Gyr$) and a higher recycled fraction. This naturally explains why PSR J0653+0443 can simultaneously exhibit both extreme eccentricity and recycled pulsar properties.

\section{Future Detection Prospects for Extremely Eccentric BNS}
\label{sec:4}

The rarity of extremely eccentric BNS systems naturally motivates an assessment of their detectability in current and future pulsar surveys. We first consider the prospects for additional discoveries within the ongoing FAST GPPS survey, which currently provides the most sensitive search for Galactic pulsars. Building on the Poisson expectation value $\lambda_{\rm GPPS}$ derived in Section~\ref{sec:3}, {we estimate the conditional probability that the completed GPPS sample contains two or more extremely eccentric systems, given that at least one is present, as}
\begin{equation}
P(N_{\rm GPPS}^{e} \ge 2 \mid N_{\rm GPPS}^{e} \ge 1) = 1-\frac{\lambda_{\rm GPPS}e^{-\lambda_{\rm GPPS}}} {1-e^{-\lambda_{\rm GPPS}}}.
\end{equation}
We find that {this probability is $\sim87.5\%$, $59.2\%$, and
$14.7\%$} for systems with $e>0.90$, $0.95$, and $0.99$,
respectively. {These results indicate that, conditioned on the
presence of one such system, additional moderately eccentric systems
are plausible within the completed GPPS sample.} For the most extreme
population with $e>0.99$, however, the probability of an additional
system remains modest, suggesting that it may be challenging for FAST GPPS survey to discover another BNS system with extremely high eccentricity like PSR J0653+0443. 

Future radio facilities with substantially improved sensitivity and survey speed are expected to significantly enhance the detectability of rare BNS populations. In particular, SKA \citep[e.g.,][]{2025OJAp....854256K} is anticipated to expand the accessible survey volume and improve sensitivity to faint pulsars, thereby increasing the number of detectable Galactic BNS systems. {Adopting the SKA survey settings from \citet{2025OJAp....854256K} and under the same population model $M_1$, we obtain illustrative estimates of }$\lambda_{\rm SKA} \sim 10$, $5$, and $1$ for eccentricity thresholds of $e>0.90$, $0.95$, and $0.99$, respectively. These expectations imply that next-generation radio surveys will likely uncover a sequence of extremely eccentric systems, even with extreme eccentricity like PSR J0653+0443, enabling a more detailed population-level constraints on the formation and evolution of BNSs.

In addition to radio surveys, future space-based GW detectors operating in the sub-mHz band, like LISAmax \citep[e.g.,][]{2023CQGra..40s5022M}, eASTROD \citep[e.g.,][]{2013IJMPD..2241004N} and Folkner \citep[e.g.,][]{Mueller+2019BAAS} are expected to provide an independent probe of extremely eccentric BNS populations. As estimated in \citet{2026arXiv260316386C}, such missions may detect of order $\sim 20$ systems with $e>0.9$, forming a complementary sample that does not rely on radio emission and therefore the spin evolution. The combination of GW and radio observations may therefore offer a powerful probe to constrain the spin evolution and early dynamical history of BNS systems.

\section{Conclusions}
\label{sec:5}

In this Letter, we investigate the statistical properties and evolutionary origin of extremely eccentric BNS systems in the context of the recent discovery of PSR J0653+0443 by FAST GPPS survey. By performing compact binary population synthesis calculations that incorporate self-consistent neutron star spin evolution and realistic observational selection effects, we obtain the following main conclusions.

\begin{itemize}
\item {Extremely eccentric BNS systems are rare among the GPPS-detectable
population. BNSs with orbital eccentricity $e>0.99$ constitute only
a small fraction ($\sim0.64\%$) of detectable systems in the FAST
survey}, indicating that the discovery of PSR J0653+0443 is
statistically uncommon but still compatible with the standard binary
evolution scenario.

\item The formation of extremely eccentric BNSs is strongly linked to the second supernova event. BNS systems with $ e\gtrsim 0.9$ are preferentially produced through CCSNe, reflecting the increasing importance of large natal kicks in generating extreme orbital configurations. In addition,  systems with spin period $P<0.1s$ such as PSR J0653+0443 may have larger probability to experience substantial mass transfer  prior to the second supernova and have a larger formation time. 

\item Future surveys will play a key role in establishing the population-level properties of extremely eccentric BNS systems. {While multiple systems with moderately high eccentricities
$e\gtrsim0.9$ are plausible within the completed modelled GPPS
sample, the probability of an additional system with an eccentricity
comparable to PSR J0653+0443 ($e\gtrsim0.99$) remains modest.} In contrast, next-generation radio telescopes like SKA and sub-mHz GW detectors are expected to substantially increase these extremely eccentric systems, enabling tighter constraints on neutron star formation channels and spin evolution.
\end{itemize}

Overall, the discovery of PSR J0653+0443 provides a valuable probe of the immediate post-supernova phase of binary evolution. Continued radio observations, together with future multi-messenger detections, will offer an increasingly powerful means to test the physics of neutron star formation and binary interaction in the high-eccentricity regime.

\section*{Acknowledgements}
We would like to express our gratitude to Professor JinLin Han and Dr. Zonglin Yang at NAOC for generously sharing the parameters of PSR J0653+0443 and for helpful discussions. This work is partly supported by the National Natural Science Foundation of China (grant nos. 12533009, and 12273050), the Strategic Priority Program of the Chinese Academy of Sciences (grant no. XDB0550300), the National Astronomical Observatory of China (grant no. E4TG660101), and the Postdoctoral Fellowship Program of CPSF under Grant Number GZB20250735 (ZC).

\section*{Data Availability}
The data underlying this article will be shared on reasonable request to the corresponding author.

\bibliographystyle{mnras}
\bibliography{ref}

@ARTICLE{2023CQGra..40s5022M,
       author = {{Martens}, W. and {Khan}, M. and {Bayle}, J.-B.},
        title = "{LISAmax: improving the low-frequency gravitational-wave sensitivity by two orders of magnitude}",
      journal = {Classical and Quantum Gravity},
         year = 2023,
        month = oct,
       volume = {40},
       number = {19},
          eid = {195022},
        pages = {195022},
          doi = {10.1088/1361-6382/acf3c7},
archivePrefix = {arXiv},
       eprint = {2304.08287},
 primaryClass = {gr-qc},
       adsurl = {https://ui.adsabs.harvard.edu/abs/2023CQGra..40s5022M}
}

@ARTICLE{2013IJMPD..2241004N,
       author = {{Ni}, Wei-Tou},
        title = "{Astrod-Gw Overview and Progress}",
      journal = {International Journal of Modern Physics D},
         year = 2013,
        month = jan,
       volume = {22},
       number = {1},
          eid = {1341004},
        pages = {1341004},
          doi = {10.1142/S0218271813410046},
archivePrefix = {arXiv},
       eprint = {1212.2816},
 primaryClass = {astro-ph.IM},
       adsurl = {https://ui.adsabs.harvard.edu/abs/2013IJMPD..2241004N}
}

@INPROCEEDINGS{Mueller+2019BAAS,
       author = {{Mueller}, Guido and {Baker}, John and {Barke}, Simon and {Bender}, Peter L. and {Berti}, Emanuele and {Caldwell}, Robert and {Conklin}, John W. and {Cornish}, Neil and {Ferrara}, Elizabeth C. and {Holley-Bockelmann}, Kelly and {Kamai}, Brittany and {Larson}, Shane L. and {Livas}, Jeff and {McWilliams}, Sean T. and {Mueller}, Guido and {Natarajan}, Priyamvada and {Rioux}, Norman and {Sankar}, Shannon S. and {Schnittman}, Jeremy and {Shoemaker}, Deirdre and {Slutsky}, Jacob and {Stebbins}, Robin and {Thorpe}, Ira and {Ziemer}, John},
        title = "{Space based gravitational wave astronomy beyond LISA}",
    booktitle = {Bulletin of the American Astronomical Society},
         year = 2019,
       volume = {51},
        month = sep,
          eid = {243},
        pages = {243},
          doi = {10.48550/arXiv.1907.11305},
archivePrefix = {arXiv},
       eprint = {1907.11305},
 primaryClass = {astro-ph.IM},
       adsurl = {https://ui.adsabs.harvard.edu/abs/2019BAAS...51g.243M}
}

@ARTICLE{2025OJAp....854256K,
       author = {{Keane}, Evan F. and {Graber}, Vanessa and {Levin}, Lina and {Tan}, Chia Min and {Johnson}, Owen A. and {Ng}, Cherry and {Pardo-Araujo}, Celsa and {Ronchi}, Michele and {Vohl}, Dany and {Xue}, Mengyao},
        title = "{A Square Kilometre Array Pulsar Census}",
      journal = {The Open Journal of Astrophysics},
         year = 2025,
        month = dec,
       volume = {8},
        pages = {54256},
          doi = {10.33232/001c.154256},
archivePrefix = {arXiv},
       eprint = {2512.16153},
 primaryClass = {astro-ph.HE},
       adsurl = {https://ui.adsabs.harvard.edu/abs/2025OJAp....854256K}
}

@ARTICLE{2026arXiv260316386C,
       author = {{Chen}, Zhiwei and {Lu}, Youjun and {Luo}, Yuchao and {Zhang}, Jihui and {Guo}, Xiao and {Liu}, Jifeng and {Ni}, Wei-Tou},
        title = "{Detectability of Nearby Binary Neutron Stars with Future sub-mHz Gravitational Wave Missions}",
      journal = {arXiv e-prints},
         year = 2026,
        month = mar,
          eid = {arXiv:2603.16386},
        pages = {arXiv:2603.16386},
          doi = {10.48550/arXiv.2603.16386},
archivePrefix = {arXiv},
       eprint = {2603.16386},
 primaryClass = {astro-ph.HE},
       adsurl = {https://ui.adsabs.harvard.edu/abs/2026arXiv260316386C}
}

@ARTICLE{2023MNRAS.526.2210S,
       author = {{Sgalletta}, Cecilia and {Iorio}, Giuliano and {Mapelli}, Michela and {Artale}, M. Celeste and {Boco}, Lumen and {Chattopadhyay}, Debatri and {Lapi}, Andrea and {Possenti}, Andrea and {Rinaldi}, Stefano and {Spera}, Mario},
        title = "{Binary neutron star populations in the Milky Way}",
      journal = {\mnras},
         year = 2023,
        month = dec,
       volume = {526},
       number = {2},
        pages = {2210-2229},
          doi = {10.1093/mnras/stad2768},
archivePrefix = {arXiv},
       eprint = {2305.04955},
 primaryClass = {astro-ph.HE},
       adsurl = {https://ui.adsabs.harvard.edu/abs/2023MNRAS.526.2210S}
}

@ARTICLE{2025arXiv250614519J,
       author = {{Jing}, W.~C. and {Han}, J.~L. and {Wang}, C. and {Wang}, P.~F. and {Wang}, T. and {Cai}, N.~N. and {Xu}, J. and {Yang}, Z.~L. and {Zhou}, D.~J. and {Yan}, Yi and {Su}, W.~Q. and {Gao}, X.~Y. and {Xie}, L.},
        title = "{FAST Pulsar Database: II. Scattering profiles of 122 Pulsars}",
      journal = {arXiv e-prints},
         year = 2025,
        month = jun,
          eid = {arXiv:2506.14519},
        pages = {arXiv:2506.14519},
          doi = {10.48550/arXiv.2506.14519},
archivePrefix = {arXiv},
       eprint = {2506.14519},
 primaryClass = {astro-ph.HE},
       adsurl = {https://ui.adsabs.harvard.edu/abs/2025arXiv250614519J}
}

@ARTICLE{2025RAA....25a4002Y,
       author = {{Yang}, Z.~L. and {Han}, J.~L. and {Wang}, T. and {Wang}, P.~F. and {Su}, W.~Q. and {Chen}, W.~C. and {Wang}, C. and {Zhou}, D.~J. and {Yan}, Y. and {Jing}, W.~C. and {Cai}, N.~N. and {Xie}, L. and {Xu}, J. and {Wang}, H.~G. and {Xu}, R.~X.},
        title = "{The FAST Galactic Plane Pulsar Snapshot Survey. VII. Six Millisecond Pulsars in Compact Orbits with Massive White Dwarf Companions}",
      journal = {Research in Astronomy and Astrophysics},
         year = 2025,
        month = jan,
       volume = {25},
       number = {1},
          eid = {014002},
        pages = {014002},
          doi = {10.1088/1674-4527/ada3b5},
archivePrefix = {arXiv},
       eprint = {2412.03063},
 primaryClass = {astro-ph.HE},
       adsurl = {https://ui.adsabs.harvard.edu/abs/2025RAA....25a4002Y}
}

@ARTICLE{2025RAA....25a4001H,
       author = {{Han}, J.~L. and {Zhou}, D.~J. and {Wang}, C. and {Su}, W.~Q. and {Yan}, Yi and {Jing}, W.~C. and {Yang}, Z.~L. and {Wang}, P.~F. and {Wang}, T. and {Xu}, J. and {Cai}, N.~N. and {Sun}, J.~H. and {Yang}, Q.~L. and {Xu}, R.~X. and {Wang}, H.~G. and {You}, X.~P.},
        title = "{The FAST Galactic Plane Pulsar Snapshot Survey. VI. The Discovery of 473 New Pulsars}",
      journal = {Research in Astronomy and Astrophysics},
         year = 2025,
        month = jan,
       volume = {25},
       number = {1},
          eid = {014001},
        pages = {014001},
          doi = {10.1088/1674-4527/ada3b7},
archivePrefix = {arXiv},
       eprint = {2411.15961},
 primaryClass = {astro-ph.HE},
       adsurl = {https://ui.adsabs.harvard.edu/abs/2025RAA....25a4001H}
}

@ARTICLE{2023MNRAS.526.2645S,
       author = {{Su}, W.~Q. and {Han}, J.~L. and {Wang}, P.~F. and {Yuan}, J.~P. and {Wang}, Chen and {Zhou}, D.~J. and {Wang}, Tao and {Yan}, Yi and {Jing}, W.~C. and {Yang}, Z.~L. and {Cai}, N.~N. and {Chen}, Xue and {Xu}, Jun and {Xie}, Lang and {Wang}, H.~G. and {Xu}, R.~X. and {You}, X.~P.},
        title = "{The FAST Galactic Plane Pulsar Snapshot Survey - III. Timing results of 30 newly discovered pulsars}",
      journal = {\mnras},
         year = 2023,
        month = dec,
       volume = {526},
       number = {2},
        pages = {2645-2656},
          doi = {10.1093/mnras/stad2159},
archivePrefix = {arXiv},
       eprint = {2305.16754},
 primaryClass = {astro-ph.HE},
       adsurl = {https://ui.adsabs.harvard.edu/abs/2023MNRAS.526.2645S}
}

@ARTICLE{2019MNRAS.482.3249Z,
       author = {{Zhu}, W.~W. and {Desvignes}, G. and {Wex}, N. and {Caballero}, R.~N. and {Champion}, D.~J. and {Demorest}, P.~B. and {Ellis}, J.~A. and {Janssen}, G.~H. and {Kramer}, M. and {Krieger}, A. and {Lentati}, L. and {Nice}, D.~J. and {Ransom}, S.~M. and {Stairs}, I.~H. and {Stappers}, B.~W. and {Verbiest}, J.~P.~W. and {Arzoumanian}, Z. and {Bassa}, C.~G. and {Burgay}, M. and {Cognard}, I. and {Crowter}, K. and {Dolch}, T. and {Ferdman}, R.~D. and {Fonseca}, E. and {Gonzalez}, M.~E. and {Graikou}, E. and {Guillemot}, L. and {Hessels}, J.~W.~T. and {Jessner}, A. and {Jones}, G. and {Jones}, M.~L. and {Jordan}, C. and {Karuppusamy}, R. and {Lam}, M.~T. and {Lazaridis}, K. and {Lazarus}, P. and {Lee}, K.~J. and {Levin}, L. and {Liu}, K. and {Lyne}, A.~G. and {McKee}, J.~W. and {McLaughlin}, M.~A. and {Os{\l}owski}, S. and {Pennucci}, T. and {Perrodin}, D. and {Possenti}, A. and {Sanidas}, S. and {Shaifullah}, G. and {Smits}, R. and {Stovall}, K. and {Swiggum}, J. and {Theureau}, G. and {Tiburzi}, C.},
        title = "{Tests of gravitational symmetries with pulsar binary J1713+0747}",
      journal = {\mnras},
         year = 2019,
        month = jan,
       volume = {482},
       number = {3},
        pages = {3249-3260},
          doi = {10.1093/mnras/sty2905},
archivePrefix = {arXiv},
       eprint = {1802.09206},
 primaryClass = {astro-ph.HE},
       adsurl = {https://ui.adsabs.harvard.edu/abs/2019MNRAS.482.3249Z}
}

@ARTICLE{2016ApJ...829...55W,
       author = {{Weisberg}, J.~M. and {Huang}, Y.},
        title = "{Relativistic Measurements from Timing the Binary Pulsar PSR B1913+16}",
      journal = {\apj},
         year = 2016,
        month = sep,
       volume = {829},
       number = {1},
          eid = {55},
        pages = {55},
          doi = {10.3847/0004-637X/829/1/55},
archivePrefix = {arXiv},
       eprint = {1606.02744},
 primaryClass = {astro-ph.HE},
       adsurl = {https://ui.adsabs.harvard.edu/abs/2016ApJ...829...55W}
}

@ARTICLE{2025arXiv251012506M,
       author = {{Meng}, Lingqi and {Freire}, Paulo C.~C. and {Stovall}, Kevin and {Wex}, Norbert and {Miao}, Xueli and {Zhu}, Weiwei and {Kramer}, Michael and {Cordes}, James M. and {Hu}, Huanchen and {Jiang}, Jinchen and {Parent}, Emilie and {Shao}, Lijing and {Stairs}, Ingrid H. and {Xue}, Mengyao and {Brazier}, Adam and {Camilo}, Fernando and {Champion}, David J. and {Chatterjee}, Shami and {Crawford}, Fronefield and {Fang}, Ziyao and {Fu}, Qiuyang and {Guo}, Yanjun and {Hessels}, Jason W.~T. and {MacLaughlin}, Maura and {Miao}, Chenchen and {Niu}, Jiarui and {Wu}, Ziwei and {Yao}, Jumei and {Yuan}, Mao and {Yue}, Youlin and {Zhang}, Chengmin},
        title = "{The double neutron star PSR J1946+2052 I. Masses and tests of general relativity}",
      journal = {arXiv e-prints},
         year = 2025,
        month = oct,
          eid = {arXiv:2510.12506},
        pages = {arXiv:2510.12506},
          doi = {10.48550/arXiv.2510.12506},
archivePrefix = {arXiv},
       eprint = {2510.12506},
 primaryClass = {astro-ph.HE},
       adsurl = {https://ui.adsabs.harvard.edu/abs/2025arXiv251012506M}
}

@ARTICLE{2024MNRAS.527.1101G,
       author = {{Gaspari}, Nicola and {Levan}, Andrew J. and {Chrimes}, Ashley A. and {Nelemans}, Gijs},
        title = "{The Galactic neutron star population - II. Systemic velocities and merger locations of binary neutron stars}",
      journal = {\mnras},
         year = 2024,
        month = jan,
       volume = {527},
       number = {1},
        pages = {1101-1113},
          doi = {10.1093/mnras/stad3259},
archivePrefix = {arXiv},
       eprint = {2310.14773},
 primaryClass = {astro-ph.HE},
       adsurl = {https://ui.adsabs.harvard.edu/abs/2024MNRAS.527.1101G}
}

@ARTICLE{1982ApJ...253..908T,
       author = {{Taylor}, J.~H. and {Weisberg}, J.~M.},
        title = "{A new test of general relativity - Gravitational radiation and the binary pulsar PSR 1913+16}",
      journal = {\apj},
         year = 1982,
        month = feb,
       volume = {253},
        pages = {908-920},
          doi = {10.1086/159690},
       adsurl = {https://ui.adsabs.harvard.edu/abs/1982ApJ...253..908T}
}

@ARTICLE{2013A&A...549A.137I,
       author = {{Irrgang}, A. and {Wilcox}, B. and {Tucker}, E. and {Schiefelbein}, L.},
        title = "{Milky Way mass models for orbit calculations}",
      journal = {\aap},
         year = 2013,
        month = jan,
       volume = {549},
          eid = {A137},
        pages = {A137},
          doi = {10.1051/0004-6361/201220540},
archivePrefix = {arXiv},
       eprint = {1211.4353},
 primaryClass = {astro-ph.GA},
       adsurl = {https://ui.adsabs.harvard.edu/abs/2013A&A...549A.137I}
}

@ARTICLE{2004ApJ...605..759B,
       author = {{Bhat}, N.~D. Ramesh and {Cordes}, James M. and {Camilo}, Fernando and {Nice}, David J. and {Lorimer}, Duncan R.},
        title = "{Multifrequency Observations of Radio Pulse Broadening and Constraints on Interstellar Electron Density Microstructure}",
      journal = {\apj},
         year = 2004,
        month = apr,
       volume = {605},
       number = {2},
        pages = {759-783},
          doi = {10.1086/382680},
archivePrefix = {arXiv},
       eprint = {astro-ph/0401067},
 primaryClass = {astro-ph},
       adsurl = {https://ui.adsabs.harvard.edu/abs/2004ApJ...605..759B}
}

@ARTICLE{2014ApJ...784...59S,
       author = {{Szary}, Andrzej and {Zhang}, Bing and {Melikidze}, George I. and {Gil}, Janusz and {Xu}, Ren-Xin},
        title = "{Radio Efficiency of Pulsars}",
      journal = {\apj},
         year = 2014,
        month = mar,
       volume = {784},
       number = {1},
          eid = {59},
        pages = {59},
          doi = {10.1088/0004-637X/784/1/59},
archivePrefix = {arXiv},
       eprint = {1402.0228},
 primaryClass = {astro-ph.HE},
       adsurl = {https://ui.adsabs.harvard.edu/abs/2014ApJ...784...59S}
}

@ARTICLE{1998MNRAS.298..625T,
       author = {{Tauris}, T.~M. and {Manchester}, R.~N.},
        title = "{On the Evolution of Pulsar Beams}",
      journal = {\mnras},
         year = 1998,
        month = aug,
       volume = {298},
       number = {3},
        pages = {625-636},
          doi = {10.1046/j.1365-8711.1998.01369.x},
       adsurl = {https://ui.adsabs.harvard.edu/abs/1998MNRAS.298..625T}
}

@ARTICLE{2002astro.ph..7156C,
       author = {{Cordes}, J.~M. and {Lazio}, T.~J.~W.},
        title = "{NE2001.I. A New Model for the Galactic Distribution of Free Electrons and its Fluctuations}",
      journal = {arXiv e-prints},
         year = 2002,
        month = jul,
          eid = {astro-ph/0207156},
        pages = {astro-ph/0207156},
          doi = {10.48550/arXiv.astro-ph/0207156},
archivePrefix = {arXiv},
       eprint = {astro-ph/0207156},
 primaryClass = {astro-ph},
       adsurl = {https://ui.adsabs.harvard.edu/abs/2002astro.ph..7156C}
}

@ARTICLE{2025ApJ...980..181C,
       author = {{Chu}, Qingbo and {Lu}, Youjun and {Yu}, Shenghua},
        title = "{Spin Evolution and Mass Distribution of Galactic Binary Neutron Stars}",
      journal = {\apj},
         year = 2025,
        month = feb,
       volume = {980},
       number = {2},
          eid = {181},
        pages = {181},
          doi = {10.3847/1538-4357/ad90e2},
archivePrefix = {arXiv},
       eprint = {2412.15464},
 primaryClass = {astro-ph.HE},
       adsurl = {https://ui.adsabs.harvard.edu/abs/2025ApJ...980..181C}
}

@ARTICLE{1975ApJ...195L..51H,
       author = {{Hulse}, R.~A. and {Taylor}, J.~H.},
        title = "{Discovery of a pulsar in a binary system.}",
      journal = {\apjl},
         year = 1975,
        month = jan,
       volume = {195},
        pages = {L51-L53},
          doi = {10.1086/181708},
       adsurl = {https://ui.adsabs.harvard.edu/abs/1975ApJ...195L..51H}
}

@ARTICLE{2011IJMPD..20..989N,
       author = {{Nan}, Rendong and {Li}, Di and {Jin}, Chengjin and {Wang}, Qiming and {Zhu}, Lichun and {Zhu}, Wenbai and {Zhang}, Haiyan and {Yue}, Youling and {Qian}, Lei},
        title = "{The Five-Hundred Aperture Spherical Radio Telescope (fast) Project}",
      journal = {International Journal of Modern Physics D},
         year = 2011,
        month = jan,
       volume = {20},
       number = {6},
        pages = {989-1024},
          doi = {10.1142/S0218271811019335},
archivePrefix = {arXiv},
       eprint = {1105.3794},
 primaryClass = {astro-ph.IM},
       adsurl = {https://ui.adsabs.harvard.edu/abs/2011IJMPD..20..989N}
}

@ARTICLE{2021RAA....21..107H,
       author = {{Han}, J.~L. and {Wang}, Chen and {Wang}, P.~F. and {Wang}, Tao and {Zhou}, D.~J. and {Sun}, Jing-Hai and {Yan}, Yi and {Su}, Wei-Qi and {Jing}, Wei-Cong and {Chen}, Xue and {Gao}, X.~Y. and {Hou}, Li-Gang and {Xu}, Jun and {Lee}, K.~J. and {Wang}, Na and {Jiang}, Peng and {Xu}, Ren-Xin and {Yan}, Jun and {Gan}, Heng-Qian and {Guan}, Xin and {Huang}, Wen-Jun and {Jiang}, Jin-Chen and {Li}, Hui and {Men}, Yun-Peng and {Sun}, Chun and {Wang}, Bo-Jun and {Wang}, H.~G. and {Wang}, Shuang-Qiang and {Xie}, Jin-Tao and {Xu}, Heng and {Yao}, Rui and {You}, Xiao-Peng and {Yu}, D.~J. and {Yuan}, Jian-Ping and {Yuen}, Rai and {Zhang}, Chun-Feng and {Zhu}, Yan},
        title = "{The FAST Galactic Plane Pulsar Snapshot survey: I. Project design and pulsar discoveries}",
      journal = {Research in Astronomy and Astrophysics},
         year = 2021,
        month = jun,
       volume = {21},
       number = {5},
          eid = {107},
        pages = {107},
          doi = {10.1088/1674-4527/21/5/107},
archivePrefix = {arXiv},
       eprint = {2105.08460},
 primaryClass = {astro-ph.HE},
       adsurl = {https://ui.adsabs.harvard.edu/abs/2021RAA....21..107H}
}

@ARTICLE{2024MNRAS.530.1506S,
       author = {{Su}, W.~Q. and {Han}, J.~L. and {Yang}, Z.~L. and {Wang}, P.~F. and {Yuan}, J.~P. and {Wang}, C. and {Zhou}, D.~J. and {Wang}, T. and {Yan}, Y. and {Jing}, W.~C. and {Cai}, N.~N. and {Xie}, L. and {Xu}, J. and {Wang}, H.~G. and {Xu}, R.~X. and {You}, X.~P.},
        title = "{The FAST Galactic Plane Pulsar Snapshot Survey - V. PSR J1901+0658 in a double neutron star system}",
      journal = {\mnras},
         year = 2024,
        month = may,
       volume = {530},
       number = {2},
        pages = {1506-1511},
          doi = {10.1093/mnras/stae888},
archivePrefix = {arXiv},
       eprint = {2403.11635},
 primaryClass = {astro-ph.HE},
       adsurl = {https://ui.adsabs.harvard.edu/abs/2024MNRAS.530.1506S}
}

@ARTICLE{2025RAA....25a4003W,
       author = {{Wang}, P.~F. and {Han}, J.~L. and {Yang}, Z.~L. and {Wang}, T. and {Wang}, C. and {Su}, W.~Q. and {Xu}, J. and {Zhou}, D.~J. and {Yan}, Yi and {Jing}, W.~C. and {Cai}, N.~N. and {Yuan}, J.~P. and {Xu}, R.~X. and {Wang}, H.~G. and {You}, X.~P.},
        title = "{The FAST Galactic Plane Pulsar Snapshot Survey. VIII. 116 Binary Pulsars}",
      journal = {Research in Astronomy and Astrophysics},
         year = 2025,
        month = jan,
       volume = {25},
       number = {1},
          eid = {014003},
        pages = {014003},
          doi = {10.1088/1674-4527/ada3b8},
archivePrefix = {arXiv},
       eprint = {2412.03062},
 primaryClass = {astro-ph.HE},
       adsurl = {https://ui.adsabs.harvard.edu/abs/2025RAA....25a4003W}
}

@ARTICLE{2022MNRAS.509.1557C,
       author = {{Chu}, Qingbo and {Yu}, Shenghua and {Lu}, Youjun},
        title = "{Formation and evolution of binary neutron stars: mergers and their host galaxies}",
      journal = {\mnras},
         year = 2022,
        month = jan,
       volume = {509},
       number = {2},
        pages = {1557-1586},
          doi = {10.1093/mnras/stab2882},
archivePrefix = {arXiv},
       eprint = {2110.04687},
 primaryClass = {astro-ph.GA},
       adsurl = {https://ui.adsabs.harvard.edu/abs/2022MNRAS.509.1557C}
}

@ARTICLE{2001MNRAS.322..231K,
       author = {{Kroupa}, Pavel},
        title = "{On the variation of the initial mass function}",
      journal = {\mnras},
         year = 2001,
        month = apr,
       volume = {322},
       number = {2},
        pages = {231-246},
          doi = {10.1046/j.1365-8711.2001.04022.x},
archivePrefix = {arXiv},
       eprint = {astro-ph/0009005},
 primaryClass = {astro-ph},
       adsurl = {https://ui.adsabs.harvard.edu/abs/2001MNRAS.322..231K}
}

@ARTICLE{1989ApJ...347..998E,
       author = {{Eggleton}, Peter P. and {Fitchett}, Michael J. and {Tout}, Christopher A.},
        title = "{The Distribution of Visual Binaries with Two Bright Components}",
      journal = {\apj},
         year = 1989,
        month = dec,
       volume = {347},
        pages = {998},
          doi = {10.1086/168190},
       adsurl = {https://ui.adsabs.harvard.edu/abs/1989ApJ...347..998E}
}

@ARTICLE{1998MNRAS.296.1019H,
       author = {{Han}, Zhanwen},
        title = "{The formation of double degenerates and related objects}",
      journal = {\mnras},
         year = 1998,
        month = jun,
       volume = {296},
       number = {4},
        pages = {1019-1040},
          doi = {10.1046/j.1365-8711.1998.01475.x},
       adsurl = {https://ui.adsabs.harvard.edu/abs/1998MNRAS.296.1019H}
}

@ARTICLE{2001A&A...380..151D,
       author = {{Douchin}, F. and {Haensel}, P.},
        title = "{A unified equation of state of dense matter and neutron star structure}",
      journal = {\aap},
         year = 2001,
        month = dec,
       volume = {380},
        pages = {151-167},
          doi = {10.1051/0004-6361:20011402},
archivePrefix = {arXiv},
       eprint = {astro-ph/0111092},
 primaryClass = {astro-ph},
       adsurl = {https://ui.adsabs.harvard.edu/abs/2001A&A...380..151D}
}

@ARTICLE{2006ApJ...643..332F,
       author = {{Faucher-Gigu{\`e}re}, Claude-Andr{\'e} and {Kaspi}, Victoria M.},
        title = "{Birth and Evolution of Isolated Radio Pulsars}",
      journal = {\apj},
         year = 2006,
        month = may,
       volume = {643},
       number = {1},
        pages = {332-355},
          doi = {10.1086/501516},
archivePrefix = {arXiv},
       eprint = {astro-ph/0512585},
 primaryClass = {astro-ph},
       adsurl = {https://ui.adsabs.harvard.edu/abs/2006ApJ...643..332F}
}

@ARTICLE{2014ApJS..212....6O,
       author = {{Olausen}, S.~A. and {Kaspi}, V.~M.},
        title = "{The McGill Magnetar Catalog}",
      journal = {\apjs},
         year = 2014,
        month = may,
       volume = {212},
       number = {1},
          eid = {6},
        pages = {6},
          doi = {10.1088/0067-0049/212/1/6},
archivePrefix = {arXiv},
       eprint = {1309.4167},
 primaryClass = {astro-ph.HE},
       adsurl = {https://ui.adsabs.harvard.edu/abs/2014ApJS..212....6O}
}

@ARTICLE{1984ApJ...277..355W,
       author = {{Webbink}, R.~F.},
        title = "{Double white dwarfs as progenitors of R Coronae Borealis stars and type I supernovae.}",
      journal = {\apj},
         year = 1984,
        month = feb,
       volume = {277},
        pages = {355-360},
          doi = {10.1086/161701},
       adsurl = {https://ui.adsabs.harvard.edu/abs/1984ApJ...277..355W}
}

@ARTICLE{2021A&A...650A.107M,
       author = {{Marchant}, Pablo and {Pappas}, Kaliro{\"e} M.~W. and {Gallegos-Garcia}, Monica and {Berry}, Christopher P.~L. and {Taam}, Ronald E. and {Kalogera}, Vicky and {Podsiadlowski}, Philipp},
        title = "{The role of mass transfer and common envelope evolution in the formation of merging binary black holes}",
      journal = {\aap},
         year = 2021,
        month = jun,
       volume = {650},
          eid = {A107},
        pages = {A107},
          doi = {10.1051/0004-6361/202039992},
archivePrefix = {arXiv},
       eprint = {2103.09243},
 primaryClass = {astro-ph.SR},
       adsurl = {https://ui.adsabs.harvard.edu/abs/2021A&A...650A.107M}
}

@ARTICLE{2023LRCA....9....2R,
       author = {{R{\"o}pke}, Friedrich K. and {De Marco}, Orsola},
        title = "{Simulations of common-envelope evolution in binary stellar systems: physical models and numerical techniques}",
      journal = {Living Reviews in Computational Astrophysics},
         year = 2023,
        month = dec,
       volume = {9},
       number = {1},
          eid = {2},
        pages = {2},
          doi = {10.1007/s41115-023-00017-x},
archivePrefix = {arXiv},
       eprint = {2212.07308},
 primaryClass = {astro-ph.SR},
       adsurl = {https://ui.adsabs.harvard.edu/abs/2023LRCA....9....2R}
}

@ARTICLE{1980PASJ...32..303M,
       author = {{Miyaji}, S. and {Nomoto}, K. and {Yokoi}, K. and {Sugimoto}, D.},
        title = "{Supernova triggered by electron captures.}",
      journal = {\pasj},
         year = 1980,
        month = jan,
       volume = {32},
        pages = {303-329},
       adsurl = {https://ui.adsabs.harvard.edu/abs/1980PASJ...32..303M}
}

@ARTICLE{2015MNRAS.451.2123T,
       author = {{Tauris}, Thomas M. and {Langer}, Norbert and {Podsiadlowski}, Philipp},
        title = "{Ultra-stripped supernovae: progenitors and fate}",
      journal = {\mnras},
         year = 2015,
        month = aug,
       volume = {451},
       number = {2},
        pages = {2123-2144},
          doi = {10.1093/mnras/stv990},
archivePrefix = {arXiv},
       eprint = {1505.00270},
 primaryClass = {astro-ph.SR},
       adsurl = {https://ui.adsabs.harvard.edu/abs/2015MNRAS.451.2123T}
}

@ARTICLE{2002MNRAS.329..897H,
       author = {{Hurley}, Jarrod R. and {Tout}, Christopher A. and {Pols}, Onno R.},
        title = "{Evolution of binary stars and the effect of tides on binary populations}",
      journal = {\mnras},
         year = 2002,
        month = feb,
       volume = {329},
       number = {4},
        pages = {897-928},
          doi = {10.1046/j.1365-8711.2002.05038.x},
archivePrefix = {arXiv},
       eprint = {astro-ph/0201220},
 primaryClass = {astro-ph},
       adsurl = {https://ui.adsabs.harvard.edu/abs/2002MNRAS.329..897H}
}

@ARTICLE{2018MNRAS.481.4009V,
       author = {{Vigna-G{\'o}mez}, Alejandro and {Neijssel}, Coenraad J. and {Stevenson}, Simon and {Barrett}, Jim W. and {Belczynski}, Krzysztof and {Justham}, Stephen and {de Mink}, Selma E. and {M{\"u}ller}, Bernhard and {Podsiadlowski}, Philipp and {Renzo}, Mathieu and {Sz{\'e}csi}, Dorottya and {Mandel}, Ilya},
        title = "{On the formation history of Galactic double neutron stars}",
      journal = {\mnras},
         year = 2018,
        month = dec,
       volume = {481},
       number = {3},
        pages = {4009-4029},
          doi = {10.1093/mnras/sty2463},
archivePrefix = {arXiv},
       eprint = {1805.07974},
 primaryClass = {astro-ph.SR},
       adsurl = {https://ui.adsabs.harvard.edu/abs/2018MNRAS.481.4009V}
}

@ARTICLE{2010MNRAS.404.1081R,
       author = {{Ridley}, J.~P. and {Lorimer}, D.~R.},
        title = "{Isolated pulsar spin evolution on the diagram}",
      journal = {\mnras},
         year = 2010,
        month = may,
       volume = {404},
       number = {2},
        pages = {1081-1088},
          doi = {10.1111/j.1365-2966.2010.16342.x},
archivePrefix = {arXiv},
       eprint = {1001.2483},
 primaryClass = {astro-ph.GA},
       adsurl = {https://ui.adsabs.harvard.edu/abs/2010MNRAS.404.1081R}
}

@ARTICLE{1964PhRv..136.1224P,
       author = {{Peters}, P.~C.},
        title = "{Gravitational Radiation and the Motion of Two Point Masses}",
      journal = {Physical Review},
         year = 1964,
        month = nov,
       volume = {136},
       number = {4B},
        pages = {1224-1232},
          doi = {10.1103/PhysRev.136.B1224},
       adsurl = {https://ui.adsabs.harvard.edu/abs/1964PhRv..136.1224P}
}

@ARTICLE{2020MNRAS.494.1587C,
       author = {{Chattopadhyay}, Debatri and {Stevenson}, Simon and {Hurley}, Jarrod R. and {Rossi}, Luca J. and {Flynn}, Chris},
        title = "{Modelling double neutron stars: radio and gravitational waves}",
      journal = {\mnras},
         year = 2020,
        month = may,
       volume = {494},
       number = {2},
        pages = {1587-1610},
          doi = {10.1093/mnras/staa756},
archivePrefix = {arXiv},
       eprint = {1912.02415},
 primaryClass = {astro-ph.HE},
       adsurl = {https://ui.adsabs.harvard.edu/abs/2020MNRAS.494.1587C}
}

@ARTICLE{2015ApJS..216...29B,
       author = {{Bovy}, Jo},
        title = "{galpy: A python Library for Galactic Dynamics}",
      journal = {\apjs},
         year = 2015,
        month = feb,
       volume = {216},
       number = {2},
          eid = {29},
        pages = {29},
          doi = {10.1088/0067-0049/216/2/29},
archivePrefix = {arXiv},
       eprint = {1412.3451},
 primaryClass = {astro-ph.GA},
       adsurl = {https://ui.adsabs.harvard.edu/abs/2015ApJS..216...29B}
}

@ARTICLE{1991PhR...203....1B,
       author = {{Bhattacharya}, D. and {van den Heuvel}, E.~P.~J.},
        title = "{Formation and evolution of binary and millisecond radio pulsars}",
      journal = {\physrep},
         year = 1991,
        month = jan,
       volume = {203},
       number = {1-2},
        pages = {1-124},
          doi = {10.1016/0370-1573(91)90064-S},
       adsurl = {https://ui.adsabs.harvard.edu/abs/1991PhR...203....1B}
}

@ARTICLE{2003Natur.426..531B,
       author = {{Burgay}, M. and {D'Amico}, N. and {Possenti}, A. and {Manchester}, R.~N. and {Lyne}, A.~G. and {Joshi}, B.~C. and {McLaughlin}, M.~A. and {Kramer}, M. and {Sarkissian}, J.~M. and {Camilo}, F. and {Kalogera}, V. and {Kim}, C. and {Lorimer}, D.~R.},
        title = "{An increased estimate of the merger rate of double neutron stars from observations of a highly relativistic system}",
      journal = {\nat},
         year = 2003,
        month = dec,
       volume = {426},
       number = {6966},
        pages = {531-533},
          doi = {10.1038/nature02124},
archivePrefix = {arXiv},
       eprint = {astro-ph/0312071},
 primaryClass = {astro-ph},
       adsurl = {https://ui.adsabs.harvard.edu/abs/2003Natur.426..531B}
}

@ARTICLE{1985ApJ...294L..25D,
       author = {{Dewey}, R.~J. and {Taylor}, J.~H. and {Weisberg}, J.~M. and {Stokes}, G.~H.},
        title = "{A search for low-luminosity pulsars.}",
      journal = {\apjl},
         year = 1985,
        month = jul,
       volume = {294},
        pages = {L25-L29},
          doi = {10.1086/184502},
       adsurl = {https://ui.adsabs.harvard.edu/abs/1985ApJ...294L..25D}
}

@ARTICLE{1997ApJS..113..367B,
       author = {{Bildsten}, Lars and {Chakrabarty}, Deepto and {Chiu}, John and {Finger}, Mark H. and {Koh}, Danny T. and {Nelson}, Robert W. and {Prince}, Thomas A. and {Rubin}, Bradley C. and {Scott}, D. Matthew and {Stollberg}, Mark and {Vaughan}, Brian A. and {Wilson}, Colleen A. and {Wilson}, Robert B.},
        title = "{Observations of Accreting Pulsars}",
      journal = {\apjs},
         year = 1997,
        month = dec,
       volume = {113},
       number = {2},
        pages = {367-408},
          doi = {10.1086/313060},
archivePrefix = {arXiv},
       eprint = {astro-ph/9707125},
 primaryClass = {astro-ph},
       adsurl = {https://ui.adsabs.harvard.edu/abs/1997ApJS..113..367B}
}

@ARTICLE{2019MNRAS.486.3213A,
       author = {{Andrews}, Jeff J. and {Zezas}, Andreas},
        title = "{Double neutron star formation: merger times, systemic velocities, and travel distances}",
      journal = {\mnras},
         year = 2019,
        month = jul,
       volume = {486},
       number = {3},
        pages = {3213-3227},
          doi = {10.1093/mnras/stz1066},
archivePrefix = {arXiv},
       eprint = {1904.06137},
 primaryClass = {astro-ph.HE},
       adsurl = {https://ui.adsabs.harvard.edu/abs/2019MNRAS.486.3213A}
}

@ARTICLE{2014ARA&A..52..415M,
       author = {{Madau}, Piero and {Dickinson}, Mark},
        title = "{Cosmic Star-Formation History}",
      journal = {\araa},
         year = 2014,
        month = aug,
       volume = {52},
        pages = {415-486},
          doi = {10.1146/annurev-astro-081811-125615},
archivePrefix = {arXiv},
       eprint = {1403.0007},
 primaryClass = {astro-ph.CO},
       adsurl = {https://ui.adsabs.harvard.edu/abs/2014ARA&A..52..415M}
}

@ARTICLE{2016Natur.534..512B,
       author = {{Belczynski}, Krzysztof and {Holz}, Daniel E. and {Bulik}, Tomasz and {O'Shaughnessy}, Richard},
        title = "{The first gravitational-wave source from the isolated evolution of two stars in the 40-100 solar mass range}",
      journal = {\nat},
         year = 2016,
        month = jun,
       volume = {534},
       number = {7608},
        pages = {512-515},
          doi = {10.1038/nature18322},
archivePrefix = {arXiv},
       eprint = {1602.04531},
 primaryClass = {astro-ph.HE},
       adsurl = {https://ui.adsabs.harvard.edu/abs/2016Natur.534..512B}
}

@ARTICLE{2018MNRAS.480.2011G,
       author = {{Giacobbo}, Nicola and {Mapelli}, Michela},
        title = "{The progenitors of compact-object binaries: impact of metallicity, common envelope and natal kicks}",
      journal = {\mnras},
         year = 2018,
        month = oct,
       volume = {480},
       number = {2},
        pages = {2011-2030},
          doi = {10.1093/mnras/sty1999},
archivePrefix = {arXiv},
       eprint = {1806.00001},
 primaryClass = {astro-ph.HE},
       adsurl = {https://ui.adsabs.harvard.edu/abs/2018MNRAS.480.2011G}
}

@ARTICLE{2016ApJ...819..108B,
       author = {{Belczynski}, Krzysztof and {Repetto}, Serena and {Holz}, Daniel E. and {O'Shaughnessy}, Richard and {Bulik}, Tomasz and {Berti}, Emanuele and {Fryer}, Christopher and {Dominik}, Michal},
        title = "{Compact Binary Merger Rates: Comparison with LIGO/Virgo Upper Limits}",
      journal = {\apj},
         year = 2016,
        month = mar,
       volume = {819},
       number = {2},
          eid = {108},
        pages = {108},
          doi = {10.3847/0004-637X/819/2/108},
archivePrefix = {arXiv},
       eprint = {1510.04615},
 primaryClass = {astro-ph.HE},
       adsurl = {https://ui.adsabs.harvard.edu/abs/2016ApJ...819..108B}
}

@ARTICLE{2019ApJ...880L...8A,
       author = {{Andrews}, Jeff J. and {Mandel}, Ilya},
        title = "{Double Neutron Star Populations and Formation Channels}",
      journal = {\apjl},
         year = 2019,
        month = jul,
       volume = {880},
       number = {1},
          eid = {L8},
        pages = {L8},
          doi = {10.3847/2041-8213/ab2ed1},
archivePrefix = {arXiv},
       eprint = {1904.12745},
 primaryClass = {astro-ph.HE},
       adsurl = {https://ui.adsabs.harvard.edu/abs/2019ApJ...880L...8A}
}

@ARTICLE{2022LRR....25....1M,
       author = {{Mandel}, Ilya and {Broekgaarden}, Floor S.},
        title = "{Rates of compact object coalescences}",
      journal = {Living Reviews in Relativity},
         year = 2022,
        month = dec,
       volume = {25},
       number = {1},
          eid = {1},
        pages = {1},
          doi = {10.1007/s41114-021-00034-3},
archivePrefix = {arXiv},
       eprint = {2107.14239},
 primaryClass = {astro-ph.HE},
       adsurl = {https://ui.adsabs.harvard.edu/abs/2022LRR....25....1M}
}

@ARTICLE{2019MNRAS.488...47F,
       author = {{Fragione}, Giacomo and {Grishin}, Evgeni and {Leigh}, Nathan W.~C. and {Perets}, Hagai B. and {Perna}, Rosalba},
        title = "{Black hole and neutron star mergers in galactic nuclei}",
      journal = {\mnras},
         year = 2019,
        month = sep,
       volume = {488},
       number = {1},
        pages = {47-63},
          doi = {10.1093/mnras/stz1651},
archivePrefix = {arXiv},
       eprint = {1811.10627},
 primaryClass = {astro-ph.GA},
       adsurl = {https://ui.adsabs.harvard.edu/abs/2019MNRAS.488...47F}
}

@ARTICLE{Stovall+2018ApJ...854L..22S,
       author = {{Stovall}, K. and {Freire}, P.~C.~C. and {Chatterjee}, S. and {Demorest}, P.~B. and {Lorimer}, D.~R. and {McLaughlin}, M.~A. and {Pol}, N. and {van Leeuwen}, J. and {Wharton}, R.~S. and {Allen}, B. and {Boyce}, M. and {Brazier}, A. and {Caballero}, K. and {Camilo}, F. and {Camuccio}, R. and {Cordes}, J.~M. and {Crawford}, F. and {Deneva}, J.~S. and {Ferdman}, R.~D. and {Hessels}, J.~W.~T. and {Jenet}, F.~A. and {Kaspi}, V.~M. and {Knispel}, B. and {Lazarus}, P. and {Lynch}, R. and {Parent}, E. and {Patel}, C. and {Pleunis}, Z. and {Ransom}, S.~M. and {Scholz}, P. and {Seymour}, A. and {Siemens}, X. and {Stairs}, I.~H. and {Swiggum}, J. and {Zhu}, W.~W.},
        title = "{PALFA Discovery of a Highly Relativistic Double Neutron Star Binary}",
      journal = {ApJL},
         year = 2018,
        month = feb,
       volume = {854},
       number = {2},
          eid = {L22},
        pages = {L22},
          doi = {10.3847/2041-8213/aaad06},
archivePrefix = {arXiv},
       eprint = {1802.01707},
 primaryClass = {astro-ph.HE},
       adsurl = {https://ui.adsabs.harvard.edu/abs/2018ApJ...854L..22S}
}

@ARTICLE{Keith+2009MNRAS.393..623K,
       author = {{Keith}, M.~J. and {Kramer}, M. and {Lyne}, A.~G. and {Eatough}, R.~P. and {Stairs}, I.~H. and {Possenti}, A. and {Camilo}, F. and {Manchester}, R.~N.},
        title = "{PSR J1753-2240: a mildly recycled pulsar in an eccentric binary system}",
      journal = {\mnras},
         year = 2009,
        month = feb,
       volume = {393},
       number = {2},
        pages = {623-627},
          doi = {10.1111/j.1365-2966.2008.14234.x},
archivePrefix = {arXiv},
       eprint = {0811.2027},
 primaryClass = {astro-ph},
       adsurl = {https://ui.adsabs.harvard.edu/abs/2009MNRAS.393..623K}
}

@ARTICLE{2018MNRAS.476.4315N,
       author = {{Ng}, C. and {Kruckow}, M.~U. and {Tauris}, T.~M. and {Lyne}, A.~G. and {Freire}, P.~C.~C. and {Ridolfi}, A. and {Caiazzo}, I. and {Heyl}, J. and {Kramer}, M. and {Cameron}, A.~D. and {Champion}, D.~J. and {Stappers}, B.},
        title = "{PSR J1755-2550: a young radio pulsar with a massive, compact companion}",
      journal = {\mnras},
         year = 2018,
        month = jun,
       volume = {476},
       number = {4},
        pages = {4315-4326},
          doi = {10.1093/mnras/sty482},
archivePrefix = {arXiv},
       eprint = {1802.08248},
 primaryClass = {astro-ph.HE},
       adsurl = {https://ui.adsabs.harvard.edu/abs/2018MNRAS.476.4315N}
}

@ARTICLE{Swiggum+2023ApJ...944..154S,
       author = {{Swiggum}, J.~K. and {Pleunis}, Z. and {Parent}, E. and {Kaplan}, D.~L. and {McLaughlin}, M.~A. and {Stairs}, I.~H. and {Spiewak}, R. and {Agazie}, G.~Y. and {Chawla}, P. and {DeCesar}, M.~E. and {Dolch}, T. and {Fiore}, W. and {Fonseca}, E. and {Istrate}, A.~G. and {Kaspi}, V.~M. and {Kondratiev}, V.~I. and {van Leeuwen}, J. and {Levin}, L. and {Lewis}, E.~F. and {Lynch}, R.~S. and {McEwen}, A.~E. and {Al Noori}, H. and {Ransom}, S.~M. and {Siemens}, X. and {Surnis}, M.},
        title = "{The Green Bank North Celestial Cap Survey. VII. 12 New Pulsar Timing Solutions}",
      journal = {\apj},
         year = 2023,
        month = feb,
       volume = {944},
       number = {2},
          eid = {154},
        pages = {154},
          doi = {10.3847/1538-4357/acb43f},
archivePrefix = {arXiv},
       eprint = {2212.03926},
 primaryClass = {astro-ph.HE},
       adsurl = {https://ui.adsabs.harvard.edu/abs/2023ApJ...944..154S}
}

@ARTICLE{Ridolfi+2019MNRAS.490.3860R,
       author = {{Ridolfi}, A. and {Freire}, P.~C.~C. and {Gupta}, Y. and {Ransom}, S.~M.},
        title = "{Upgraded Giant Metrewave Radio Telescope timing of NGC 1851A: a possible millisecond pulsar - neutron star system}",
      journal = {\mnras},
         year = 2019,
        month = dec,
       volume = {490},
       number = {3},
        pages = {3860-3874},
          doi = {10.1093/mnras/stz2645},
archivePrefix = {arXiv},
       eprint = {1909.06163},
 primaryClass = {astro-ph.HE},
       adsurl = {https://ui.adsabs.harvard.edu/abs/2019MNRAS.490.3860R}
}

@ARTICLE{Wu+2023ApJ...958L..17W,
       author = {{Wu}, Q.~D. and {Wang}, N. and {Yuan}, J.~P. and {Li}, D. and {Wang}, P. and {Xue}, M.~Y. and {Zhu}, W.~W. and {Miao}, C.~C. and {Yan}, W.~M. and {Wang}, J.~B. and {Yao}, J.~M. and {Wang}, S.~Q. and {Sun}, S.~N. and {Kou}, F.~F. and {Zhao}, D. and {Chen}, Y.~T. and {Dang}, S.~J. and {Feng}, Y. and {Liu}, Z.~J. and {Miao}, X.~L. and {Meng}, L.~Q. and {Yuan}, M. and {Niu}, C.~H. and {Niu}, J.~R. and {Qian}, L. and {Wang}, S. and {Xie}, X.~Y. and {Xiao}, Y.~F. and {Yue}, Y.~L. and {You}, S.~P. and {Yu}, X.~H. and {Zhao}, R.~S. and {Yuen}, R. and {Zhou}, X. and {Zhang}, L.},
        title = "{PSR J2150+3427: A Possible Double Neutron Star System}",
      journal = {\apjl},
         year = 2023,
        month = nov,
       volume = {958},
       number = {1},
          eid = {L17},
        pages = {L17},
          doi = {10.3847/2041-8213/ad0887},
       adsurl = {https://ui.adsabs.harvard.edu/abs/2023ApJ...958L..17W}
}

@ARTICLE{Cameron+2023MNRAS.523.5064C,
       author = {{Cameron}, A.~D. and {Bailes}, M. and {Champion}, D.~J. and {Freire}, P.~C.~C. and {Kramer}, M. and {McLaughlin}, M.~A. and {Ng}, C. and {Possenti}, A. and {Ridolfi}, A. and {Tauris}, T.~M. and {Wahl}, H.~M. and {Wex}, N.},
        title = "{New constraints on the kinematic, relativistic, and evolutionary properties of the PSR J1757-1854 double neutron star system}",
      journal = {\mnras},
         year = 2023,
        month = aug,
       volume = {523},
       number = {4},
        pages = {5064-5085},
          doi = {10.1093/mnras/stad1712},
archivePrefix = {arXiv},
       eprint = {2305.14733},
 primaryClass = {astro-ph.HE},
       adsurl = {https://ui.adsabs.harvard.edu/abs/2023MNRAS.523.5064C}
}

@ARTICLE{Kramer+2021PhRvX..11d1050K,
       author = {{Kramer}, M. and {Stairs}, I.~H. and {Manchester}, R.~N. and {Wex}, N. and {Deller}, A.~T. and {Coles}, W.~A. and {Ali}, M. and {Burgay}, M. and {Camilo}, F. and {Cognard}, I. and {Damour}, T. and {Desvignes}, G. and {Ferdman}, R.~D. and {Freire}, P.~C.~C. and {Grondin}, S. and {Guillemot}, L. and {Hobbs}, G.~B. and {Janssen}, G. and {Karuppusamy}, R. and {Lorimer}, D.~R. and {Lyne}, A.~G. and {McKee}, J.~W. and {McLaughlin}, M. and {M{\"u}nch}, L.~E. and {Perera}, B.~B.~P. and {Pol}, N. and {Possenti}, A. and {Sarkissian}, J. and {Stappers}, B.~W. and {Theureau}, G.},
        title = "{Strong-Field Gravity Tests with the Double Pulsar}",
      journal = {PhRvX},
         year = 2021,
        month = oct,
       volume = {11},
       number = {4},
          eid = {041050},
        pages = {041050},
          doi = {10.1103/PhysRevX.11.041050},
archivePrefix = {arXiv},
       eprint = {2112.06795},
 primaryClass = {astro-ph.HE},
       adsurl = {https://ui.adsabs.harvard.edu/abs/2021PhRvX..11d1050K}
}

@ARTICLE{Martinez+2015ApJ...812..143M,
       author = {{Martinez}, J.~G. and {Stovall}, K. and {Freire}, P.~C.~C. and {Deneva}, J.~S. and {Jenet}, F.~A. and {McLaughlin}, M.~A. and {Bagchi}, M. and {Bates}, S.~D. and {Ridolfi}, A.},
        title = "{Pulsar J0453+1559: A Double Neutron Star System with a Large Mass Asymmetry}",
      journal = {\apj},
         year = 2015,
        month = oct,
       volume = {812},
       number = {2},
          eid = {143},
        pages = {143},
          doi = {10.1088/0004-637X/812/2/143},
archivePrefix = {arXiv},
       eprint = {1509.08805},
 primaryClass = {astro-ph.HE},
       adsurl = {https://ui.adsabs.harvard.edu/abs/2015ApJ...812..143M}
}

@ARTICLE{Leeuwen+2015ApJ...798..118V,
       author = {{van Leeuwen}, J. and {Kasian}, L. and {Stairs}, I.~H. and {Lorimer}, D.~R. and {Camilo}, F. and {Chatterjee}, S. and {Cognard}, I. and {Desvignes}, G. and {Freire}, P.~C.~C. and {Janssen}, G.~H. and {Kramer}, M. and {Lyne}, A.~G. and {Nice}, D.~J. and {Ransom}, S.~M. and {Stappers}, B.~W. and {Weisberg}, J.~M.},
        title = "{The Binary Companion of Young, Relativistic Pulsar J1906+0746}",
      journal = {\apj},
         year = 2015,
        month = jan,
       volume = {798},
       number = {2},
          eid = {118},
        pages = {118},
          doi = {10.1088/0004-637X/798/2/118},
archivePrefix = {arXiv},
       eprint = {1411.1518},
 primaryClass = {astro-ph.SR},
       adsurl = {https://ui.adsabs.harvard.edu/abs/2015ApJ...798..118V}
}

@ARTICLE{McEwen+2024ApJ...962..167M,
       author = {{McEwen}, A.~E. and {Swiggum}, J.~K. and {Kaplan}, D.~L. and {Tan}, C.~M. and {Meyers}, B.~W. and {Fonseca}, E. and {Agazie}, G.~Y. and {Chawla}, P. and {Crowter}, K. and {DeCesar}, M.~E. and {Dolch}, T. and {Dong}, F.~A. and {Fiore}, W. and {Fonseca}, E. and {Good}, D.~C. and {Istrate}, A.~G. and {Kaspi}, V.~M. and {Kondratiev}, V.~I. and {van Leeuwen}, J. and {Levin}, L. and {Lewis}, E.~F. and {Lynch}, R.~S. and {Masui}, K.~W. and {McKee}, J.~W. and {McLaughlin}, M.~A. and {Al Noori}, H. and {Parent}, E. and {Ransom}, S.~M. and {Siemens}, X. and {Spiewak}, R. and {Stairs}, I.~H.},
        title = "{The Green Bank North Celestial Cap Survey. IX. Timing Follow-up for 128 Pulsars}",
      journal = {\apj},
         year = 2024,
        month = feb,
       volume = {962},
       number = {2},
          eid = {167},
        pages = {167},
          doi = {10.3847/1538-4357/ad11f0},
archivePrefix = {arXiv},
       eprint = {2312.07471},
 primaryClass = {astro-ph.HE},
       adsurl = {https://ui.adsabs.harvard.edu/abs/2024ApJ...962..167M}
}

@ARTICLE{Sengar+2022MNRAS.512.5782S,
       author = {{Sengar}, R. and {Balakrishnan}, V. and {Stevenson}, S. and {Bailes}, M. and {Barr}, E.~D. and {Bhat}, N.~D.~R. and {Burgay}, M. and {Bernadich}, M.~C. i. and {Cameron}, A.~D. and {Champion}, D.~J. and {Chen}, W. and {Flynn}, C.~M.~L. and {Jameson}, A. and {Johnston}, S. and {Keith}, M.~J. and {Kramer}, M. and {Morello}, V. and {Ng}, C. and {Possenti}, A. and {Stappers}, B. and {Shannon}, R.~M. and {van Straten}, W. and {Wongphechauxsorn}, J.},
        title = "{The High Time Resolution Universe Pulsar Survey - XVII. PSR J1325-6253, a low eccentricity double neutron star system from an ultra-stripped supernova}",
      journal = {\mnras},
         year = 2022,
        month = jun,
       volume = {512},
       number = {4},
        pages = {5782-5792},
          doi = {10.1093/mnras/stac821},
archivePrefix = {arXiv},
       eprint = {2204.06869},
 primaryClass = {astro-ph.HE},
       adsurl = {https://ui.adsabs.harvard.edu/abs/2022MNRAS.512.5782S}
}

@ARTICLE{Janssen+2008AA...490..753J,
       author = {{Janssen}, G.~H. and {Stappers}, B.~W. and {Kramer}, M. and {Nice}, D.~J. and {Jessner}, A. and {Cognard}, I. and {Purver}, M.~B.},
        title = "{Multi-telescope timing of PSR J1518+4904}",
      journal = {\aap},
         year = 2008,
        month = nov,
       volume = {490},
       number = {2},
        pages = {753-761},
          doi = {10.1051/0004-6361:200810076},
archivePrefix = {arXiv},
       eprint = {0808.2292},
 primaryClass = {astro-ph},
       adsurl = {https://ui.adsabs.harvard.edu/abs/2008A&A...490..753J}
}

@ARTICLE{Weisberg+2016ApJ...829...55W,
       author = {{Weisberg}, J.~M. and {Huang}, Y.},
        title = "{Relativistic Measurements from Timing the Binary Pulsar PSR B1913+16}",
      journal = {\apj},
         year = 2016,
        month = sep,
       volume = {829},
       number = {1},
          eid = {55},
        pages = {55},
          doi = {10.3847/0004-637X/829/1/55},
archivePrefix = {arXiv},
       eprint = {1606.02744},
 primaryClass = {astro-ph.HE},
       adsurl = {https://ui.adsabs.harvard.edu/abs/2016ApJ...829...55W}
}

@ARTICLE{Swiggum+2015ApJ...805..156S,
       author = {{Swiggum}, J.~K. and {Rosen}, R. and {McLaughlin}, M.~A. and {Lorimer}, D.~R. and {Heatherly}, S. and {Lynch}, R. and {Scoles}, S. and {Hockett}, T. and {Filik}, E. and {Marlowe}, J.~A. and {Barlow}, B.~N. and {Weaver}, M. and {Hilzendeger}, M. and {Ernst}, S. and {Crowley}, R. and {Stone}, E. and {Miller}, B. and {Nunez}, R. and {Trevino}, G. and {Doehler}, M. and {Cramer}, A. and {Yencsik}, D. and {Thorley}, J. and {Andrews}, R. and {Laws}, A. and {Wenger}, K. and {Teter}, L. and {Snyder}, T. and {Dittmann}, A. and {Gray}, S. and {Carter}, M. and {McGough}, C. and {Dydiw}, S. and {Pruett}, C. and {Fink}, J. and {Vanderhout}, A.},
        title = "{PSR J1930-1852: a Pulsar in the Widest Known Orbit around Another Neutron Star}",
      journal = {\apj},
         year = 2015,
        month = jun,
       volume = {805},
       number = {2},
          eid = {156},
        pages = {156},
          doi = {10.1088/0004-637X/805/2/156},
archivePrefix = {arXiv},
       eprint = {1503.06276},
 primaryClass = {astro-ph.HE},
       adsurl = {https://ui.adsabs.harvard.edu/abs/2015ApJ...805..156S}
}

@ARTICLE{Lazarus+2016ApJ...831..150L,
       author = {{Lazarus}, P. and {Freire}, P.~C.~C. and {Allen}, B. and {Aulbert}, C. and {Bock}, O. and {Bogdanov}, S. and {Brazier}, A. and {Camilo}, F. and {Cardoso}, F. and {Chatterjee}, S. and {Cordes}, J.~M. and {Crawford}, F. and {Deneva}, J.~S. and {Eggenstein}, H. -B. and {Fehrmann}, H. and {Ferdman}, R. and {Hessels}, J.~W.~T. and {Jenet}, F.~A. and {Karako-Argaman}, C. and {Kaspi}, V.~M. and {Knispel}, B. and {Lynch}, R. and {van Leeuwen}, J. and {Machenschalk}, B. and {Madsen}, E. and {McLaughlin}, M.~A. and {Patel}, C. and {Ransom}, S.~M. and {Scholz}, P. and {Seymour}, A. and {Siemens}, X. and {Spitler}, L.~G. and {Stairs}, I.~H. and {Stovall}, K. and {Swiggum}, J. and {Venkataraman}, A. and {Zhu}, W.~W.},
        title = "{Einstein@Home Discovery of a Double Neutron Star Binary in the PALFA Survey}",
      journal = {\apj},
         year = 2016,
        month = nov,
       volume = {831},
       number = {2},
          eid = {150},
        pages = {150},
          doi = {10.3847/0004-637X/831/2/150},
archivePrefix = {arXiv},
       eprint = {1608.08211},
 primaryClass = {astro-ph.HE},
       adsurl = {https://ui.adsabs.harvard.edu/abs/2016ApJ...831..150L}
}

@ARTICLE{Haniewicz+2021MNRAS.500.4620H,
       author = {{Haniewicz}, H.~T. and {Ferdman}, R.~D. and {Freire}, P.~C.~C. and {Champion}, D.~J. and {Bunting}, K.~A. and {Lorimer}, D.~R. and {McLaughlin}, M.~A.},
        title = "{Precise mass measurements for the double neutron star system J1829+2456}",
      journal = {\mnras},
         year = 2021,
        month = feb,
       volume = {500},
       number = {4},
        pages = {4620-4627},
          doi = {10.1093/mnras/staa3466},
archivePrefix = {arXiv},
       eprint = {2007.07565},
 primaryClass = {astro-ph.SR},
       adsurl = {https://ui.adsabs.harvard.edu/abs/2021MNRAS.500.4620H}
}

@ARTICLE{Ferdman+2014MNRAS.443.2183F,
       author = {{Ferdman}, R.~D. and {Stairs}, I.~H. and {Kramer}, M. and {Janssen}, G.~H. and {Bassa}, C.~G. and {Stappers}, B.~W. and {Demorest}, P.~B. and {Cognard}, I. and {Desvignes}, G. and {Theureau}, G. and {Burgay}, M. and {Lyne}, A.~G. and {Manchester}, R.~N. and {Possenti}, A.},
        title = "{PSR J1756-2251: a pulsar with a low-mass neutron star companion}",
      journal = {\mnras},
         year = 2014,
        month = sep,
       volume = {443},
       number = {3},
        pages = {2183-2196},
          doi = {10.1093/mnras/stu1223},
archivePrefix = {arXiv},
       eprint = {1406.5507},
 primaryClass = {astro-ph.SR},
       adsurl = {https://ui.adsabs.harvard.edu/abs/2014MNRAS.443.2183F}
}

@ARTICLE{Fonseca+2014ApJ...787...82F,
       author = {{Fonseca}, Emmanuel and {Stairs}, Ingrid H. and {Thorsett}, Stephen E.},
        title = "{A Comprehensive Study of Relativistic Gravity Using PSR B1534+12}",
      journal = {\apj},
         year = 2014,
        month = may,
       volume = {787},
       number = {1},
          eid = {82},
        pages = {82},
          doi = {10.1088/0004-637X/787/1/82},
archivePrefix = {arXiv},
       eprint = {1402.4836},
 primaryClass = {astro-ph.HE},
       adsurl = {https://ui.adsabs.harvard.edu/abs/2014ApJ...787...82F}
}

@ARTICLE{Martinez+2017ApJ...851L..29M,
       author = {{Martinez}, J.~G. and {Stovall}, K. and {Freire}, P.~C.~C. and {Deneva}, J.~S. and {Tauris}, T.~M. and {Ridolfi}, A. and {Wex}, N. and {Jenet}, F.~A. and {McLaughlin}, M.~A. and {Bagchi}, M.},
        title = "{Pulsar J1411+2551: A Low-mass Double Neutron Star System}",
      journal = {\apjl},
         year = 2017,
        month = dec,
       volume = {851},
       number = {2},
          eid = {L29},
        pages = {L29},
          doi = {10.3847/2041-8213/aa9d87},
archivePrefix = {arXiv},
       eprint = {1711.09804},
 primaryClass = {astro-ph.HE},
       adsurl = {https://ui.adsabs.harvard.edu/abs/2017ApJ...851L..29M}
}

@ARTICLE{2017MNRAS.472.2422M,
       author = {{Mapelli}, Michela and {Giacobbo}, Nicola and {Ripamonti}, Emanuele and {Spera}, Mario},
        title = "{The cosmic merger rate of stellar black hole binaries from the Illustris simulation}",
      journal = {\mnras},
         year = 2017,
        month = dec,
       volume = {472},
       number = {2},
        pages = {2422-2435},
          doi = {10.1093/mnras/stx2123},
archivePrefix = {arXiv},
       eprint = {1708.05722},
 primaryClass = {astro-ph.GA},
       adsurl = {https://ui.adsabs.harvard.edu/abs/2017MNRAS.472.2422M}
}

@ARTICLE{2012ApJ...759...52D,
       author = {{Dominik}, Michal and {Belczynski}, Krzysztof and {Fryer}, Christopher and {Holz}, Daniel E. and {Berti}, Emanuele and {Bulik}, Tomasz and {Mandel}, Ilya and {O'Shaughnessy}, Richard},
        title = "{Double Compact Objects. I. The Significance of the Common Envelope on Merger Rates}",
      journal = {\apj},
         year = 2012,
        month = nov,
       volume = {759},
       number = {1},
          eid = {52},
        pages = {52},
          doi = {10.1088/0004-637X/759/1/52},
archivePrefix = {arXiv},
       eprint = {1202.4901},
 primaryClass = {astro-ph.HE},
       adsurl = {https://ui.adsabs.harvard.edu/abs/2012ApJ...759...52D}
}

@ARTICLE{2013ApJ...779...72D,
       author = {{Dominik}, Michal and {Belczynski}, Krzysztof and {Fryer}, Christopher and {Holz}, Daniel E. and {Berti}, Emanuele and {Bulik}, Tomasz and {Mandel}, Ilya and {O'Shaughnessy}, Richard},
        title = "{Double Compact Objects. II. Cosmological Merger Rates}",
      journal = {\apj},
         year = 2013,
        month = dec,
       volume = {779},
       number = {1},
          eid = {72},
        pages = {72},
          doi = {10.1088/0004-637X/779/1/72},
archivePrefix = {arXiv},
       eprint = {1308.1546},
 primaryClass = {astro-ph.HE},
       adsurl = {https://ui.adsabs.harvard.edu/abs/2013ApJ...779...72D}
}

@ARTICLE{2024ApJ...976...24B,
       author = {{Boesky}, Adam P. and {Broekgaarden}, Floor S. and {Berger}, Edo},
        title = "{Investigating the Cosmological Rate of Compact Object Mergers from Isolated Massive Binary Stars}",
      journal = {\apj},
         year = 2024,
        month = nov,
       volume = {976},
       number = {1},
          eid = {24},
        pages = {24},
          doi = {10.3847/1538-4357/ad7fe3},
archivePrefix = {arXiv},
       eprint = {2405.01630},
 primaryClass = {astro-ph.HE},
       adsurl = {https://ui.adsabs.harvard.edu/abs/2024ApJ...976...24B}
}

@ARTICLE{2020ApJ...898..152S,
       author = {{Santoliquido}, Filippo and {Mapelli}, Michela and {Bouffanais}, Yann and {Giacobbo}, Nicola and {Di Carlo}, Ugo N. and {Rastello}, Sara and {Artale}, M. Celeste and {Ballone}, Alessandro},
        title = "{The Cosmic Merger Rate Density Evolution of Compact Binaries Formed in Young Star Clusters and in Isolated Binaries}",
      journal = {\apj},
         year = 2020,
        month = aug,
       volume = {898},
       number = {2},
          eid = {152},
        pages = {152},
          doi = {10.3847/1538-4357/ab9b78},
archivePrefix = {arXiv},
       eprint = {2004.09533},
 primaryClass = {astro-ph.HE},
       adsurl = {https://ui.adsabs.harvard.edu/abs/2020ApJ...898..152S}
}

@ARTICLE{2007ApJ...662..504B,
       author = {{Belczynski}, Krzysztof and {Taam}, Ronald E. and {Kalogera}, Vassiliki and {Rasio}, Frederic A. and {Bulik}, Tomasz},
        title = "{On the Rarity of Double Black Hole Binaries: Consequences for Gravitational Wave Detection}",
      journal = {\apj},
         year = 2007,
        month = jun,
       volume = {662},
       number = {1},
        pages = {504-511},
          doi = {10.1086/513562},
archivePrefix = {arXiv},
       eprint = {astro-ph/0612032},
 primaryClass = {astro-ph},
       adsurl = {https://ui.adsabs.harvard.edu/abs/2007ApJ...662..504B}
}

@ARTICLE{2004ApJ...601.1058I,
       author = {{Ivanova}, Natalia and {Taam}, Ronald E.},
        title = "{Thermal Timescale Mass Transfer and the Evolution of White Dwarf Binaries}",
      journal = {\apj},
         year = 2004,
        month = feb,
       volume = {601},
       number = {2},
        pages = {1058-1066},
          doi = {10.1086/380561},
archivePrefix = {arXiv},
       eprint = {astro-ph/0310126},
 primaryClass = {astro-ph},
       adsurl = {https://ui.adsabs.harvard.edu/abs/2004ApJ...601.1058I}
}

@ARTICLE{2000MNRAS.315..543H,
       author = {{Hurley}, Jarrod R. and {Pols}, Onno R. and {Tout}, Christopher A.},
        title = "{Comprehensive analytic formulae for stellar evolution as a function of mass and metallicity}",
      journal = {\mnras},
         year = 2000,
        month = jul,
       volume = {315},
       number = {3},
        pages = {543-569},
          doi = {10.1046/j.1365-8711.2000.03426.x},
archivePrefix = {arXiv},
       eprint = {astro-ph/0001295},
 primaryClass = {astro-ph},
       adsurl = {https://ui.adsabs.harvard.edu/abs/2000MNRAS.315..543H}
}

@ARTICLE{2010A&A...521A..85Y,
       author = {{Yu}, S. and {Jeffery}, C.~S.},
        title = "{The gravitational wave signal from diverse populations of double white dwarf binaries in the Galaxy}",
      journal = {\aap},
         year = 2010,
        month = oct,
       volume = {521},
          eid = {A85},
        pages = {A85},
          doi = {10.1051/0004-6361/201014827},
archivePrefix = {arXiv},
       eprint = {1007.4267},
 primaryClass = {astro-ph.SR},
       adsurl = {https://ui.adsabs.harvard.edu/abs/2010A&A...521A..85Y}
}

@ARTICLE{2005MNRAS.360..974H,
       author = {{Hobbs}, G. and {Lorimer}, D.~R. and {Lyne}, A.~G. and {Kramer}, M.},
        title = "{A statistical study of 233 pulsar proper motions}",
      journal = {\mnras},
         year = 2005,
        month = jul,
       volume = {360},
       number = {3},
        pages = {974-992},
          doi = {10.1111/j.1365-2966.2005.09087.x},
archivePrefix = {arXiv},
       eprint = {astro-ph/0504584},
 primaryClass = {astro-ph},
       adsurl = {https://ui.adsabs.harvard.edu/abs/2005MNRAS.360..974H}
}

@ARTICLE{2024ApJ...973..159C,
       author = {{Chen}, Zhiwei and {Lu}, Youjun and {Wang}, Jie and {Jiang}, Zhen and {Chu}, Qingbo and {Ma}, Xianghao},
        title = "{On the Cosmic Variance of the Merger Rate Density of Binary Neutron Stars}",
      journal = {\apj},
         year = 2024,
        month = oct,
       volume = {973},
       number = {2},
          eid = {159},
        pages = {159},
          doi = {10.3847/1538-4357/ad6de0},
archivePrefix = {arXiv},
       eprint = {2408.16968},
 primaryClass = {astro-ph.HE},
       adsurl = {https://ui.adsabs.harvard.edu/abs/2024ApJ...973..159C}
}

@ARTICLE{2025arXiv250814397C,
       author = {{Chen}, Zhiwei and {Zhang}, Jihui and {Lu}, Youjun and {Liu}, Jifeng and {Zeng}, Changwen},
        title = "{Constraining Common Envelope Evolution in Binary Neutron Star Formation with Combined Galactic and Gravitational-Wave Observations}",
      journal = {arXiv e-prints},
         year = 2025,
        month = aug,
          eid = {arXiv:2508.14397},
        pages = {arXiv:2508.14397},
          doi = {10.48550/arXiv.2508.14397},
archivePrefix = {arXiv},
       eprint = {2508.14397},
 primaryClass = {astro-ph.HE},
       adsurl = {https://ui.adsabs.harvard.edu/abs/2025arXiv250814397C}
}

@ARTICLE{2025ApJ...989L...8D,
       author = {{Disberg}, Paul and {Mandel}, Ilya},
        title = "{The Kick Velocity Distribution of Isolated Neutron Stars}",
      journal = {\apjl},
         year = 2025,
        month = aug,
       volume = {989},
       number = {1},
          eid = {L8},
        pages = {L8},
          doi = {10.3847/2041-8213/adf286},
archivePrefix = {arXiv},
       eprint = {2505.22102},
 primaryClass = {astro-ph.HE},
       adsurl = {https://ui.adsabs.harvard.edu/abs/2025ApJ...989L...8D}
}

@ARTICLE{yang_np,
       author = {{Yang}, Z.~L. and {Han}, J.~L. and {Hou}, L.~G. and {Chen}, Z.~W. and {Gong}, Y.~G. and {Guo}, Y.~L. and {Lu}, Y.~J.},
        title = "{PSR J0653+0443: An Eclipsing Millisecond Pulsar in an
Extremely Eccentric Double Neutron Star System}",
      journal = {Accepted by Nature Physics},
         year = 2026,
}

\appendix

\section{Constraint on CBPS models}

In this appendix, we introduce the procedure on constraining the model parameters in the formation and evolution of Galactic BNS systems with the CBPS code \texttt{BSE}, originally developed by \citet{2000MNRAS.315..543H,2002MNRAS.329..897H}, and later modified and extended by \citet{2010A&A...521A..85Y}, \citet{2022MNRAS.509.1557C,2025ApJ...980..181C}, and \citet{2025arXiv250814397C}. In section~\ref{app:a1}, we briefly introduce our basic CBPS model settings.  

\subsection{Basic CBPS settings}
\label{app:a1}

Within the CBPS simulation, several critical recipes are taken into consideration, including initial conditions, common envelope evolution, natal kick velocity distributions from supernova explosions, neutron star spin evolution, and gravitational wave-dominated orbital evolution. For brevity, we provide only a short introduction on these recipes, referring the reader to the original literature for comprehensive discussions \citep[e.g.,][]{2025ApJ...980..181C,2025arXiv250814397C}. 

The initial condition of the simulation on the formation and evolution of Galactic BNS systems are set as follows.  The initial mass spectrum for main-sequence (MS) binary stars are generated by assuming that, the the primary mass $M_1$ follows the initial mass function given by \citet{2001MNRAS.322..231K}, i.e., $p(M_1)\propto M_1^{-2.3}$ for $M_1>1M_{\odot}$, and the mass ratio $q=M_2/M_1$ follows a uniform distribution within $(0.01,1)$ \citep{1989ApJ...347..998E}. The initial separation $a$ between binaries is assumed to follow the distribution given by \citet{1998MNRAS.296.1019H}, 
\begin{equation}
    an(a)=
    \begin{cases}
    \alpha_{\rm sep}(\frac{a}{a_0})^{1.2} & \mathrm{for}\, a\leq a_0 \\
    \alpha_{\rm sep} & \mathrm{for}\, a_0<a<a_1,
    \end{cases}
\end{equation}
where $n(a)$ denotes for the probability of binaries with semi-major axis in the range from $a$ to $a+da$, $\alpha_{\rm sep}\sim 0.070$, $a_0=10R_{\odot}$, $a_1=5.75\times10^6 R_{\odot}$. Notably, we further assume all MS binaries are initially on circular orbits and therefore possessing initial eccentricities $e_i=0$. As for the newborn NS, we assume the equation of state to be SLy with maximum mass for non-rotating neutron stars of $M_{\rm TOV}\sim 2.06M_{\odot}$ \citep{2001A&A...380..151D}. Besides, we set the initial spin period $P_0=0.01\rm s$. The initial magnetic field follows the log-normal distribution with the mean of $13$ and the scatter of $0.55$ for NS with progenitor MS mass $\lesssim 20 M_{\odot}$ \citep[e.g.,][]{2006ApJ...643..332F}, and uniform distribution within the range of $[10^{14},10^{15}] \rm G$ for NS with progenitor MS mass $\gtrsim 20 M_{\odot}$ \citep[e.g.,][]{2014ApJS..212....6O}. 

The common envelope (CE) stage in the binary evolution is approximated by the widely-used $\alpha-$formalism \citep{1984ApJ...277..355W}, i.e., 
\begin{equation}
\frac{G(M_1-m_{1c})M_1}{\lambda_{\rm S} r_{\rm L}}=\alpha_{\rm CE}\bigg(\frac{Gm_{1c}M_2}{2a_{\rm f}}-\frac{GM_1M_2}{2a_{\rm i}}\bigg),
\end{equation}
where $\alpha_{\rm CE}$ is the CE ejection efficiency and $\lambda_{\rm S}$ is the structure parameters, which is drawn from \citet{2021A&A...650A.107M}. The symbol $r_{\rm L}$ represents the Roche lobe radius, $a_{\rm i}$ and $a_{\rm f}$ are the semi-major axis of the binary before and after the CE phase, $M_1$ and $m_{1c}$ are the masses of the star and its core, $M_2$ is the companion mass. Among systems in the CE evolution, binaries with Hertzsprung-gap (HG) donors requires special consideration due to their ambiguous core-envelope structure \citep{2004ApJ...601.1058I}. Current CBPS studies \citep[e.g.,][]{2007ApJ...662..504B,2012ApJ...759...52D, 2013ApJ...779...72D, 2018MNRAS.480.2011G, 2020ApJ...898..152S, 2024ApJ...976...24B} typically implement two contrasting approaches: the ``optimistic CE'' scenario, where all HG donors can survive the common-envelope phase, and ``pessimistic CE'' scenario, where HG donors are enforce to pre-mature merge during this phase. Notably, the merger rate density of stellar-mass black hole binaries (sBBHs) tends to favor the latter ``pessimistic CE'' prescription \citep{2017MNRAS.472.2422M}. As for BNS systems, based on Bayesian comparison approach, \citet{2025arXiv250814397C} suggest a ``median CE'' scenario with a $f_{\rm HG}$ fraction of HG donors are forced to merge in the CE stage, provide a nice agreement with both the orbital period-eccentricity distribution of Galactic BNS systems and merger rate density estimation inferred from LVK O4 run. Besides, we also note here that $\alpha_{\rm CE}<1$ implies the envelope's unbinding energy is sourced solely from orbital energy \citep[e.g.,][]{2023LRCA....9....2R}, whereas $\alpha_{\rm CE}>1$ suggests additional energy sources may contribute to envelope ejection \citep[e.g.,][]{2023MNRAS.526.2210S}. For example, \citet{2025ApJ...980..181C} \citep[see also,][]{2023MNRAS.526.2210S} demonstrated that models with $\alpha_{\rm CE}<1$ cannot reproduce the observed Galactic BNSs with long orbital period, because the orbits of their progenitors shrink significantly due to less efficient ejection of CE in these models. Therefore, in this work, we consider several CBPS models with different $f_{\rm HG}$ and $\alpha_{\rm CE}$ and those predicting local merger rate density consistent with LVK O4 constraint are listed in Table~\ref{table:bayes}.

Due to the asymmetry of the mass ejection of the supernova explosion (SNe) in the formation of NS, there may be substantial natal kick to the binary system. \citet{2005MNRAS.360..974H} first adopted the proper motion measurement of young isolated pulsars to estimate the natal kick velocity $v_{\rm k}$ and suggest a Maxwellian distribution  
\begin{equation}
\frac{dN}{Ndv_{\rm k}}=\bigg(\frac{2}{\pi}\bigg)^{\frac{1}{2}}\frac{v_{\rm k}^2}{\sigma_{\rm k}^3}{\rm exp}\bigg(-\frac{v_{\rm k}^2}{2\sigma_{\rm k}^2}\bigg),
\end{equation}
where $dN/N$ is the normalized number per kick bin $d{v_{\rm k}}$. The symbol $\sigma_{\rm k}$ represents the dispersion (also the median value) of the Maxwellian distribution. In the original paper of \citet{2005MNRAS.360..974H}, it was suggested that $\sigma_{\rm k}$ is about $\sim 265~\rm km/s$. Later, to meet with the orbital periods determined by observed Galactic BNS systems, several CBPS works \citep[e.g.,][]{2018MNRAS.481.4009V,2022MNRAS.509.1557C} extend it to a bimodal Maxwellian distribution with two different $\sigma_{\rm k,1/2}$ and their values are dependent on the SNe mechanism. Here we follow \citet{2018MNRAS.481.4009V} to set: the dispersion $\sigma_{\rm k}$ for electron-capture SNe (ECSNe) \citep[ECSNe,][]{1980PASJ...32..303M} and ultra-stripped SNe \citep[USSNe,][]{2015MNRAS.451.2123T} are assumed to be the same, i.e., $\sigma_{\rm k, 1}\sim 30\rm km s^{-1}$, while for core-collapse SNe \citep[CCSNe,][]{2002MNRAS.329..897H}, the dispersion is assumed to be $\sigma_{\rm k, 2}\sim 265\rm km s^{-1}$, following \citet{2005MNRAS.360..974H}. The orientation angles $(\omega,\phi,\beta)$ of the kick velocity are defined following the classical convention in the appendix A1 of \citet{2002MNRAS.329..897H} and are assumed to be randomly and uniformly distributed.  Here we also note that increasing the value of $\sigma_{\rm k, 1/2}$ leads to a higher fraction of extremely eccentric BNS systems; however, it simultaneously results in a lower local merger rate density and less long-orbital BNS systems, which may be incompatible with current LVK constraints and pulsar observations. In addition, we note that \cite{2025ApJ...989L...8D} recently proposed a new log-normal fit to the kick velocity distribution of isolated NSs, resulting a smaller kick compared with \citet{2005MNRAS.360..974H}. However, in this work we find that the bimodality kick distribution for CCSNe and ECSNe/USSNe is still needed, otherwise BNS systems with larger orbital period can not be efficiently produced without such a low kick component. Therefore, we test our prediction by only changing the Maxwellian distribution of kick velocity distribution for the CCSNe case to the distribution given in \cite{2025ApJ...989L...8D}, 
{
and we find that the fraction of systems with $e>0.99$ increases to $f_{>0.99}\sim3.3\%$, which remains at the percent level.
}

The spin of an NS plays a crucial role in determining its detectability as a pulsar in the FAST GPPS survey. The spin evolution of isolated NSs differs markedly from that of NSs in binary systems, because various astrophysical processes specifically in binary systems can either spin up (recycled) or spin down the NS, for example the Bondi-Hoyle-Littleton (BHL) accretion from a donor star. In this work, we follow \citet[][and references therein]{2025ApJ...980..181C} to treat the NS component formed first with a massive donor as an OB or Be-type high-mass X-ray binary (HMXB) system \citep{1997ApJS..113..367B}, depending on the accretion mechanism, i.e., Roche lobe or wind-fed accretion. Then we consider the spin evolution of the HMXB in different phases according to their detailed parameters, including the ejector phase, rapid rotator phase, propeller phase, super-Keplerian magnetic inhibition phase, sub-Keplerian magnetic inhibition phase, BHL accretion phase, subsonic settling accretion phase, subsonic propeller phase, and Roche lobe overflow phase. After the second SNe and the formation of BNS, both NS components spin down via the dipole radiation and the decay rates of the NS spin periods are given by \citep{2010MNRAS.404.1081R}, 
\begin{equation}
\dot{P}=9.73\times 10^{-16}\mu_{30}^2I_{45}^{-1}P^{-1} \rm s~s^{-1},
\end{equation}
where $\mu_{30}$ and $I_{45}$ are the magnetic dipole moment in units of $\rm 10^{30} G~cm^3$ and the moment of inertial in units of $\rm 10^{45} g~cm^2$, respectively. Here we refer readers to \citet[][see Figure 1 therein]{2025ApJ...980..181C} for our spin evolution implementation procedure.

The orbit of a BNS system will decay and circularize after its formation, due to GW radiation. The evolution of its semi-major axes $(a)$ and eccentricities $(e)$ can be estimated by \citep{1964PhRv..136.1224P}
\begin{equation}
\bigg\langle\frac{da}{dt}\bigg\rangle=-\frac{64}{5}\frac{G^3m_1m_2(m_1+m_2)}{c^5a^3(1-e^2)^{\frac{7}{2}}}\bigg(1+\frac{73}{24}e^2+\frac{37}{96}e^4\bigg),
\end{equation}
and 
\begin{equation}
\bigg\langle\frac{de}{dt}\bigg\rangle=-\frac{304}{15}e\frac{G^3m_1m_2(m_1+m_2)}{c^5a^4(1-e^2)^{\frac{5}{2}}}\bigg(1+\frac{121}{304}e^2\bigg),
\end{equation}
where $m_1$ and $m_2$ denote the primary and secondary component masses of the system.

\subsection{FAST GPPS selection effect}
\label{app:a2}

To simulate the Galactic BNS systems observed by the FAST GPPS survey, we basically follow the treatment in \citet{2020MNRAS.494.1587C} and \citet{2023MNRAS.526.2210S} to consider the localisation and radio selection effect.

\subsubsection{Localisation selection}

FAST is located at southwest of China, Guizhou province with the latitude of $\rm N ~ 25.647^{\circ}$, the longitude of $\rm E~106.856^{\circ}$, and can observe the sky area in  the declination range of $-0.9^{\circ}<\rm Dec<52.2^{\circ}$ with full gain in the zenith angle of $<26.4^{\circ}$ \citep{2011IJMPD..20..989N}. The FAST GPPS survey is designed and now working to survey the Galactic disk within the Galactic latitude of $\pm 10^{\circ}$ from the Galactic plane \citep{2021RAA....21..107H}. Therefore, only those mock pulsars located in such sky area can be detected.

To determine the final position $\vec{r_{\rm f}}$ of mock BNS systems, we model the potential of MW by three individual components, including a Plummer sphere bulge, an exponential disc, and a NFW dark matter halo, of which the fitted parameters for each component are drawn from Model III of   \citet{2013A&A...549A.137I} (see Table 3 therein for details). Then, we assign each mock BNS a velocity $\vec{v_{i}}$ by the local circular rotation curve with a random $10~\rm km/s$ dispersion.  Notably, during the second-born supernova explosion, {the natal kick will modify the initial velocity by a vector superposition $\vec{v}=\vec{v_{i}}+\vec{v}_{\rm sys,2}$, where $\vec{v}_{\rm sys,2}$ denotes the post-SN centre-of-mass recoil velocity of the surviving BNS due to the natal kick.} With these initial conditions (both positions and velocities), we simulate the dynamical evolution of BNS systems in the Galactic potential employing the \texttt{galpy} package \citep{2015ApJS..216...29B}. This allows us to track each system's trajectory from its formation epoch $t_{\rm f}$ to the present observation time, providing their final Galactic positions $\vec{r_{\rm f}}$. We only count down those located within the observable sky of FAST GPPS survey. Notably, as discussed in \citet{2020MNRAS.494.1587C}, there is small probability that the BNS system may be kicked out of the MW. Hereafter we do not take these systems into account for further calculations.

\begin{figure}
\centering
\includegraphics[width=1.0\columnwidth]{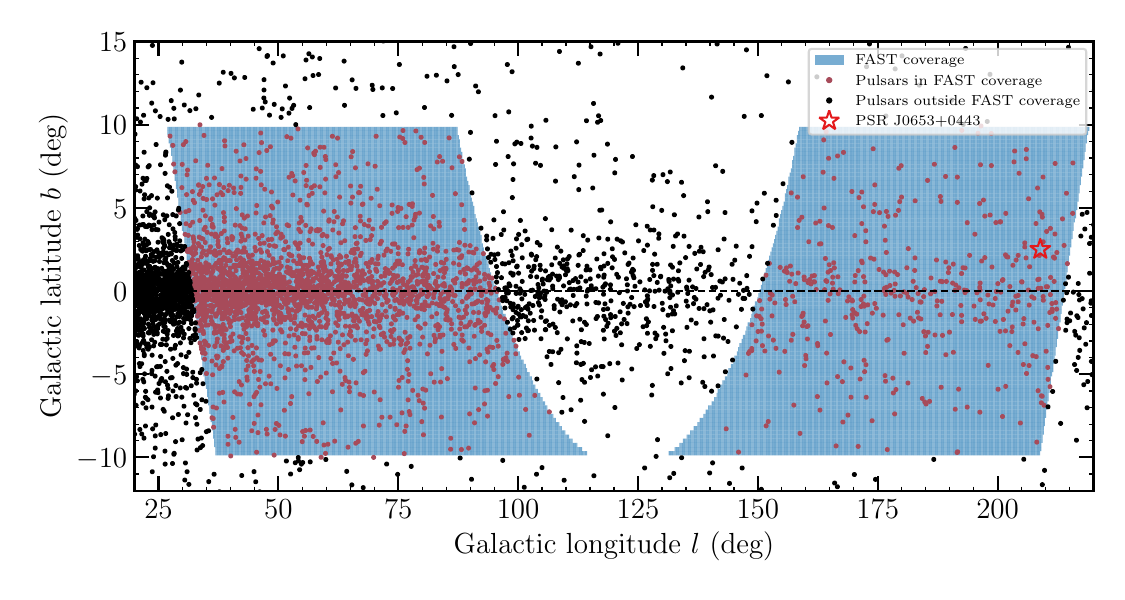}
\caption{The Sky coverage of FAST GPPS survey (blue shaded regions) in Galactic coordinates, overlaid with the spatial distribution of pulsars from our mock surviving population. Pulsars falling within the FAST GPPS survey coverage are shown as red points, while those outside the coverage are marked in black. The red star highlights the position of PSR J0653+0443. The dashed horizontal line indicates the Galactic plane ($b=0^\circ$).
}
\label{fig:f3}
\end{figure}

\subsubsection{Radio selection}

\begin{table}
\caption{ Relevant Parameter settings of FAST-GPPS survey \citep[e.g.,][]{2021RAA....21..107H,2024MNRAS.530.1506S,2025RAA....25a4003W} }
\label{table:gpps} 
\centering
\begin{tabular}{lcc} 
\hline
Parameter & Symbol & Value \\
\hline
Noise Temperature & $T_{\rm rec}+T_{\rm sky}$ & 25 K \\
Telescope Gain & $G$ & 16 K/Jy \\
Number of Polarization channel & $n_{\rm p}$ & 2\\
Integration time & $t_{\rm int}$ & 300 s\\
Band width &$\Delta f$ & 450 MHz\\ Sampling time & $\tau_{\rm samp}$ & $49.152\mu s$\\
Signal-to-noise ratio threshold & $(S/N)_{\rm min}$ & 10 \\ \hline
\end{tabular}
\end{table}

A general prerequisite for a Galactic BNS system to be observable for radio detection is that at least one NS could maintain active radio emission. In this work, we identify a BNS is radio-loud by whether the final spin of NS component generated from our spin-evolution pipeline is above the {so-called ``death line''} \citep[e.g.,][]{1991PhR...203....1B}, 
\begin{equation}
    P<P_{\rm death}=\sqrt{\frac{B}{1.7\times 10^{11}\rm G}},
\end{equation} 
where $B$ is the magnetic field of BNS. 

For a certain radio-loud BNS system to be detectable for the FAST GPPS survey, its observed flux must satisfy the threshold condition: $S\geq S_{\rm min}$. In principle, $S_{\rm min}$ is dependent on the sky location, the frequency dependence of scattering/smearing and the broadening of the beam, which can be estimated by \citep[e.g.,][]{1985ApJ...294L..25D}
\begin{equation}
S_{\rm min}=\frac{(S/N)_{\rm min}(T_{\rm rec}+T_{\rm sky})}{G\sqrt{n_{\rm p}t_{\rm int}\Delta f}}\sqrt{\frac{W_{\rm e}}{P-W_{\rm e}}},
\end{equation}
where $(S/N)_{\rm min}$ is the assumed signal-to-noise ratio for the identification of such a signal, $T_{\rm rec}$ and $T_{\rm sky}$ are respectively the receiver noise and sky background temperature, $n_{\rm p}$ is the number of polarization channels, $G$ is the gain of the telescope, $\Delta f$ is the receiver bandwidth, and $P$ is the period of the pulsar. We list the relevant settings in Table~\ref{table:gpps} drawn from the design and working group of FAST GPPS survey \citep[e.g.,][]{2021RAA....21..107H,2024MNRAS.530.1506S,2025RAA....25a4003W}. 

The effective width of a pulsar $W_{\rm e}$ can be estimated as \citep[e.g.,][]{2003Natur.426..531B}
\begin{equation}
\label{eq:we}
W_{\rm e}^2=W_i^2+\tau_{\rm samp}^2+\bigg(\tau_{\rm samp}\frac{\rm DM}{\rm DM_0}\bigg)^2+\tau_{\rm scatt}^2,
\end{equation}
where $W_i$ is the intrinsic pulse width of the pulse, $\tau_{\rm samp}$ is the sampling time, $\tau_{\rm scatt}$ is the interstellar medium (ISM) scattering time, $\rm DM$ is the dispersion measure in the direction of the pulsar, and $\rm DM_0$ is the diagonal dispersion measure of the survey. In this work, we adopt the NE2001 MW DM model from \citet{2002astro.ph..7156C} and calculate the DM for a specific BNS by its final position $\vec{r_{\rm f}}$ in MW. The ISM scatter time is approximated by the fitted results of $\tau_{\rm scatt}$ with respect to DM from \citet{2004ApJ...605..759B}. The intrinsic pulse width $W_i$ is assumed to be $0.1 P$ and declines with $P^{-1/2}$ when the period $P>10~\rm ms$, according to  \citet{2021RAA....21..107H}. 

As for estimating the intrinsic flux of a mock pulsar, we adopt the lognormal $1400\rm MHz$ radio luminosity distribution from \citet{2014ApJ...784...59S}, {i.e., ${\rm log} L_{1400}\sim N(0.5,1)$ for ${\rm log} L_{1400}\in [-3,4]$.} We then scale the luminosity to FAST's observing frequency of  $1250~\rm MHz$ assuming a power spectrum, thereby simulating the fluxes of mock BNS systems for FAST observations. {Once the radio flux of a pulsar $S_{1250}=L_{1250}/d^2_{\rm L}$} is larger than its $S_{\rm min}$, it is identified as be detectable for FAST GPPS survey. Notably, there may be significant beaming effect of the radio pulse. Here we approximate the beaming fraction $f_{\rm beaming}\sim 0.09{\rm log}(0.1P)^2+0.03 \in [0,1]$ by the results from \citet{1998MNRAS.298..625T} and conduct a further rejection sampling by $f_{\rm beaming}$ for the detection of mock BNS systems. 

\subsection{Model comparison}
\label{app:a3}

We calculate various (about $\sim 28$) CBPS models with different $f_{\rm HG}$ and $\alpha$ and estimate the local merger rate density $R_0$ for each model by \citep[e.g.,][]{2024ApJ...973..159C}
\begin{equation}
R_0=\int dt_{\rm d}P_t(t_{\rm d}) f_{\rm b}{\rm SFR}(z_{\rm b})\times N_{\rm cor}(Z\left(z_{\rm b})\right),
\label{eq:merger_rate}
\end{equation}
where $f_{\rm b}$ is the binary formation efficiency (normally $f_{\rm b}=0.5$) and $t_{\rm d}$ is the time delay of a BNS merger from its progenitor binary star formation time. The distribution $P_t(t_{\rm d})$ is directly extracted from our CBPS simulation, which is restricted within  $10\rm Myr$ to a Hubble time $t_{\rm H}$. The term ${\rm SFR}(z_{\rm b})$ is the star-formation rate density at the binary formation redshift $z_{\rm b}$ (or time $t_{\rm b}$), and $N_{\rm cor}$ is the number of merging BNSs per unit mass given a metallicity $Z(z_{\rm b})$ at $z_{\rm b}$. In this work, we draw the evolution of ${\rm SFR}(z_{\rm b})$ and $Z(z_{\rm b})$ from \citet{2014ARA&A..52..415M} and \citet{2016Natur.534..512B} respectively. Table~\ref{table:bayes} lists those models predicting local merger rate density $R_0$ within the range of $\sim 59.3_{-43.9}^{+95.4}\rm Gpc^{-3}\,yr^{-1}$ constraint by O4 run of LVK observation \citep{2026arXiv260527226T}.

For these models, we assume a constant metallicity for the MW of $Z = 0.0142$ and a uniform star formation history (SFH) similar with the treatment in \citet{2018MNRAS.481.4009V}. We simulate approximately $\sim 10^9$ main-sequence (MS) binaries (yielding $\sim 10^7$ BNS systems) per model as evolutionary templates, enabling us to quantify the relative fraction of extremely eccentric systems among all detectable BNSs. Subsequently, we apply scaling factors to boost the synthetic BNS sample size to $\sim 10^{7}$ systems and distribute them randomly throughout the MW's disk according to its density profile \citep{2013A&A...549A.137I}. 
{
Then we estimate the log-likelihood by discretizing the $(\log{P_{\rm orb}},e)$ distribution map into small pixels.
}


For each model $M_i$, we calculate
\begin{equation}
\log L_i =
\sum_j
\log\left[
\sum_{l,k}
N(\log P_{{\rm orb},j},e_j
\mid
\log P_l,e_k,
\sigma^2_{\log P_l},
\sigma^2_{e_k})
\mid_{M_i}
\right],
\end{equation}
where the inner summation is over all $(\log P_l,e_k)$ pixels, and the outer summation is over the observed Galactic field BNSs.
{
We consider nine GPPS-detectable comparison systems in total (see Table~\ref{tab:BNS}). Although not all were discovered by GPPS, these systems lie within the nominal GPPS footprint and have radio fluxes above the survey detection threshold.
}
{
To avoid using PSR J0653+0443 both in model selection and in the subsequent assessment of its rarity, our fiducial likelihood $\log L$ is calculated using the remaining eight systems. For completeness, we also calculate $\log L_*$ using all nine systems, including PSR J0653+0443.
} 
For each pixel, the 2D Gaussian contribution is weighted by the normalized number fraction of mock BNSs in that pixel, and the standard deviations are assumed to be $\sigma_{\log P_l}=0.3$ and $\sigma_{e_k}=0.03$, respectively \citep{2018MNRAS.481.4009V}. The resulting log-likelihoods are shown in Table~\ref{table:bayes}. The model M1, with $(\alpha=5,f_{\rm HG}=0.8)$, has the highest likelihoodwith or without PSR J0653+0443 and is therefore adopted as our fiducial model.

{
To assess the dependence of our main results on the choice of the fiducial population model, we repeat the GPPS detectability calculation on the fraction of highly eccentric BNSs for M1--M3, which have the three highest likelihoods among the viable models. The results are summarized in Table~\ref{tab:model_robustness}. The detectable fractions span $\sim 6.18$-$8.99\%$, $2.83$-$5.73\%$, and $0.64$-$1.14\%$ for $e>0.90$, $0.95$, and $0.99$, respectively. Although the expected counts vary more because $N_{\rm GPPS}$ is model dependent, the fraction of $e>0.99$ systems remains at the $\sim1\%$ level across M1--M3. In the fiducial M1 model, the $e>0.99$ tail contains $\sim2160$ systems out of $\sim3\times 10^5$ simulated GPPS-detectable BNSs. Thus, although this population represents only $\sim0.65\%$ of the detectable sample, the extreme-eccentricity tail is well sampled in our Monte Carlo population, and the finite-sampling uncertainty is small compared with the variation among the different population models.
}

\begin{table}
\fontsize{8}{12}\selectfont
\caption{Parameter settings for different BSE models and the corresponding log-likelihood. {Here, $\log L$ and $\log L_*$ denote the likelihoods calculated without and with PSR J0653+0443 included in the observed sample, respectively.}}
\label{table:bayes}
\centering
\begin{tabular}{lccccc}
\hline
Model & $\alpha$ & $f_{\rm HG}$ 
& $R_{0}[\rm Gpc^{-3}\,yr^{-1}]$ 
& $\log L$  & $\log L_*$ \\
\hline
M1  & 5  & 0.8 & 149 & -18.22 & -20.46 \\
M2  & 7  & 0.8 & 112 & -19.34 & -21.07 \\
M3  & 10 & 0.8 &  76 & -19.55 & -21.86\\
M4  & 10 & 0.6 & 123 & -20.16 & -22.58 \\
M5  & 3  & 1.0 & 155 & -23.54 & -25.68 \\
M6 & 5  & 1.0 &  79 & -23.04 & -25.63 \\ \hline
\end{tabular}
\end{table}

\begin{table}
\centering
\caption{{Detectable fractions and expected numbers of highly eccentric BNSs for the three highest-likelihood population models. Values in parentheses denote the corresponding expected numbers $\lambda$.}}
\label{tab:model_robustness}
\begin{tabular}{lcccc}
\hline
Model & $N_{\rm GPPS}$ &
$e>0.90$ & $e>0.95$ & $e>0.99$ \\
\hline
M1 & 48 & 6.91\% (3.32) & 3.32\% (1.59) & 0.64\% (0.31) \\
M2 & 73 & 6.18\% (4.51) & 2.83\% (2.06) & 0.77\% (0.56) \\
M3 & 92 & 8.99\% (8.27) & 5.73\% (5.27) & 1.14\% (1.05) \\
\hline
\end{tabular}
\end{table}

\section{Physical Properties of extremely Eccentric BNS systems }
\label{app:b}

We further present some supplementary information here on the statistical physical properties of extremely eccentric BNS systems detectable by FAST GPPS survey. 

\begin{figure}
\centering
\includegraphics[width=0.95\columnwidth]{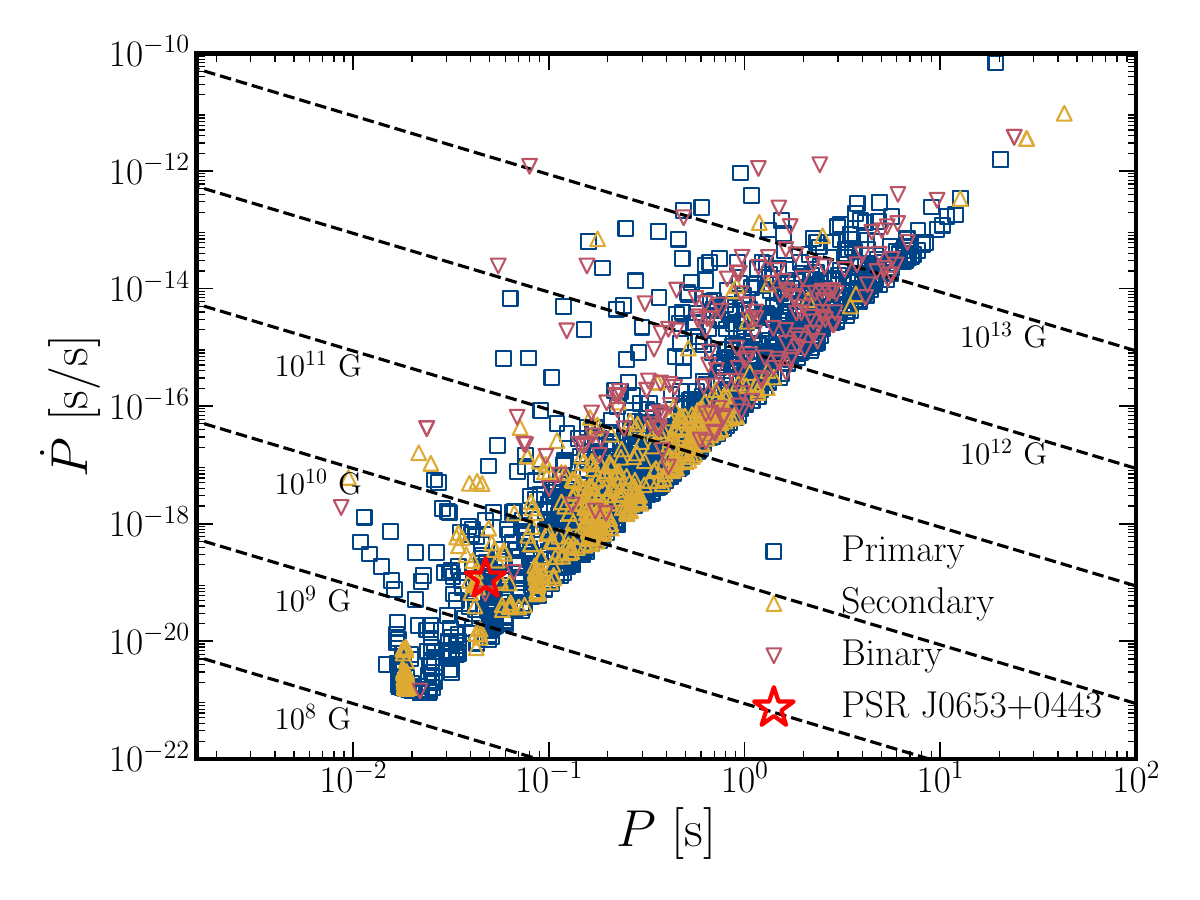}
\caption{The $P-\dot{P}$ distribution of pulsars in mock BNS systems detectable by FAST GPPS survey with $e>0.99$. The blue square, upper triangle and lower triangle represent the results for Primary-only, Secondary-only and Binary pulsar systems respectively. {Here, ``primary'' and ``secondary'' refer to the more massive and less massive NS, respectively, rather than to their birth order.} The red star marks the $P-\dot{P}$ of PSR~J0653+0443, which is a {primary-only} pulsar. The black dashed lines show the iso-magnetic levels with $B\in[10^8,10^{13}]~\rm G$. 
}
\label{fig:ppdot}
\end{figure}

Figure~\ref{fig:ppdot} shows the $P-\dot{P}$ distribution of pulsars in mock BNS systems detectable by FAST GPPS survey with $e>0.99$. Here we define : {(1) Primary-only (shown in blue squares), only the heavier NS in the BNS system is a pulsar; (2) Secondary-only (shown in yellow upper-triangles), only the lighter NS in the BNS system is a pulsar; (3) Binary (shown in magenta lower-triangles), both the NSs in the BNS system are pulsar.} As seen from this figure, while {primary-only} pulsars may possess spin $P$ in a wide range, the secondary-only pulsars are more likely to have relatively higher $P$. Here we note that the newly discovered PSR~J0653+0443 (red star in Fig.~\ref{fig:ppdot}) may be classified as a secondary-only pulsar and the existence of this extremely eccentric BNS system is compatible with the $P-\dot{P}$ distribution derived by our CBPS simulation. 

\begin{figure}
\centering
\includegraphics[width=0.95\columnwidth]{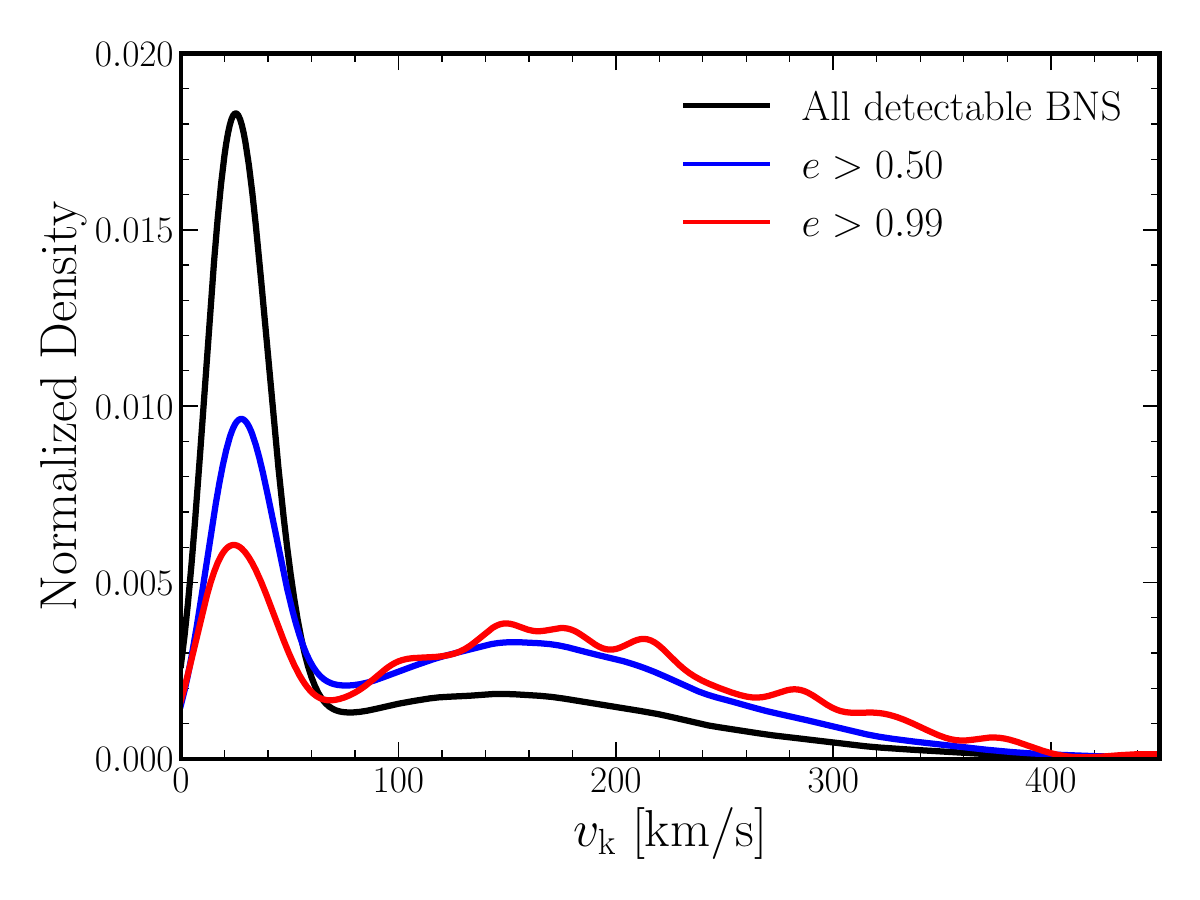}
\includegraphics[width=0.95\columnwidth]{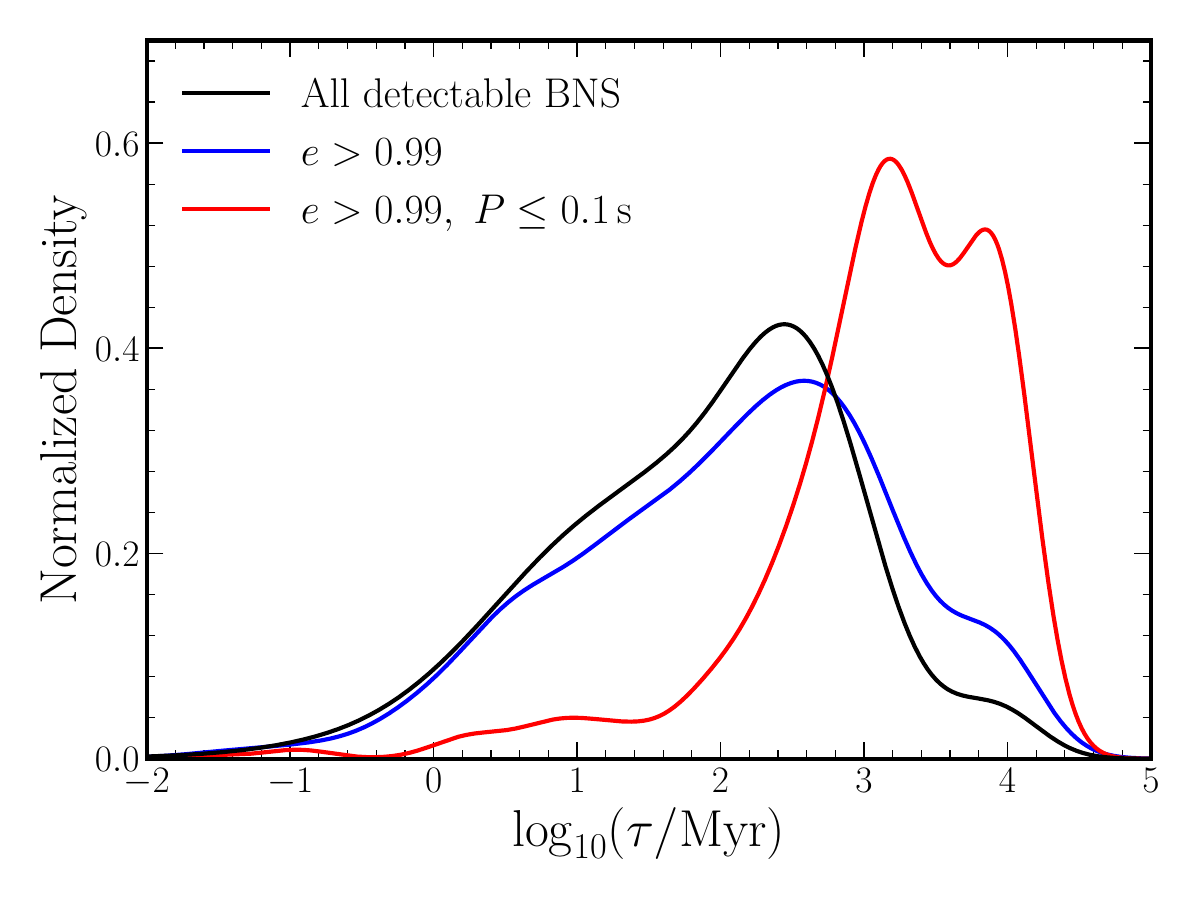}
\caption{Top panel: The normalized density distribution of kick velocity $v_{\rm k}$ of the second SNe in the formation of BNS systems. The black, blue and red lines show the results for  mock BNS systems detectable by FAST GPPS survey  with $e>0$ (all), $e>0.5$ and $e>0.99$, respectively. Bottom panel: The normalized density distribution of the formation time $\tau$ of the mock BNS systems detectable by FAST GPPS survey. The black, blue and red lines show the results of mock BNS systems with $e>0$ (all), $e>0.99$ and $e>0.99, P\leq0.1~\rm s$, respectively.
}
\label{fig:kick_tau}
\end{figure}

The top panel of Figure~\ref{fig:kick_tau} shows the normalized distribution of kick velocity $v_{\rm k}$ of the second SNe in the formation of BNS systems. It can be seen that for systems with relatively lower eccentricity threshold (black and blue), the distribution peaks at the around $\sim 50~\rm km/s$. As for extremely eccentric BNS systems with $e>0.99$ shown in red lines, the distribution exhibit a second peak with comparably height at around $\sim 150\rm km/s$, which is mainly contributed by the CCSNe channel (see Appendix~\ref{app:a1}). This result clearly indicates the critical role of CCSNe with higher natal kick velocity $v_{\rm k}$ in shaping the orbital eccentricity in the formation of BNS.  Here we also note that \citet{2026arXiv260819690D} recently showed that the broader Galactic BNS population can be reproduced with predominantly small natal kicks, with its orbital structure arising largely from remnant-mass-dependent Blaauw recoil. In their description, for an initially circular orbit with a negligible natal kick, the post-SNe eccentricity is related with the SNe ejecta mass $e\sim \Delta M/M_{\rm post}$ \citep[e.g.,][]{1961BAN....15..265B, 2026arXiv260800174G}. Considering this Blaauw recoil alone to produce a BNS system with $e>0.99$ would requires substantial SNe mass loss, thus favoring a CCSNe channel. Taken together, these results point to an important role for CCSNe channel in producing extreme eccentricities, whether through larger natal kicks or substantial SN mass loss. 

The bottom panel of Figure~\ref{fig:kick_tau} shows the normalized density distribution of the formation time $\tau$ of the mock BNS systems detectable by FAST GPPS survey. The black, blue and red lines show the results of mock BNS systems with $e>0$ (all), $e>0.99$ and $e>0.99, P\leq0.1~\rm s$ respectively. It can be seen if we restrict the extremely eccentric systems with spin period $P\leq 0.1\rm ~s$, the peak of formation $\tau$ {increases by a factor of $\sim 10$}. This can be partly explained by the observational selection effect that short-period pulsars remain detectable over longer timescales, allowing older systems to enter the observable sample. 

\begin{table*}
    \centering
    \caption{Known Galactic BNS systems in the Milky way field (i.e., not in the globular cluster).  The 2nd column shows the BNS names (candidates with $\ast$). The 3rd and 4th columns show their Galactic longitude (gl) and latitude (gb), respectively. The 5th column shows their orbital eccentricities ($e$). The references for each system are listed in the 6th column.  Galactic BNS systems with No.0-8 are located within the sky area of FAST GPPS survey originally designed in \citet{2021RAA....21..107H}. Notably, No.16 J1844-0128 is found by an extended search of GPPS survey \citep[e.g.,][]{2025RAA....25a4003W} and located outside the original sky area. Thus, we do not take this pulsar into account. In addition, pulsar J0737-3039A and B are in the same BNS as a double pulsar system denoted as {index 23}. 
    }
    \label{tab:BNS}
    \fontsize{10}{12}\selectfont %
    \begin{tabular}{lrrrrrl}
    \toprule
    No. & PSR & gl ($^{\circ}$)  & gb ($^{\circ}$)  & $e$ & Ref. \\
    \midrule
0 & J0653+0443      &  208.91 &  2.51  & 0.992387 & \cite{yang_np}   \\
1 & J1946+2052      &  57.66 &  $-$1.98  & 0.06385 & \cite{Stovall+2018ApJ...854L..22S} \\
2 & J1906+0746$^{\ast}$      &  41.60 &  0.15   & 0.08530 & \cite{Leeuwen+2015ApJ...798..118V} \\
3 & J1913+1102      &  45.25 &  0.19   & 0.08954 & \cite{Lazarus+2016ApJ...831..150L} \\
4 & J1856-0039      &  32.9 &  $-1.50$   & 0.1063 & \cite{2026arXiv260727333Y} \\
5 & J0528+3529      & 172.52  & 0.47 &0.2901& \cite{2025RAA....25a4003W} \\
6 & J1901+0658      & 40.27  & 0.98 &0.3662& \cite{2024MNRAS.530.1506S} \\
7  & J0509+3801      & 168.27   & -1.19 &0.5864& \cite{McEwen+2024ApJ...962..167M} \\
8  & B1913+16        & 49.97 & 2.12 &0.6171& \cite{Weisberg+2016ApJ...829...55W} \\ \hline
9  & J1757$-$1854    & 9.97   &2.88  &0.6058& \cite{Cameron+2023MNRAS.523.5064C} \\
10  & J2150+3427      & 85.64  & -15.00 &0.6015& \cite{Wu+2023ApJ...958L..17W} \\
11 & J1930$-$1852    & 19.98  & -16.87 &0.3989& \cite{Swiggum+2015ApJ...805..156S} \\
12 & J1759+5036$^{\ast}$      & 78.15   &28.60 & 0.3083 & \cite{McEwen+2024ApJ...962..167M} \\
13 & J1753$-$2240$^{\ast}$    & 6.30  &1.66 & 0.3036 & \cite{Keith+2009MNRAS.393..623K} \\
14 & B1534+12        & 19.85   &  48.34 &0.2737& \cite{Fonseca+2014ApJ...787...82F} \\
15 & J1518+4904      & 80.81   & 54.28 &0.2495& \cite{Janssen+2008AA...490..753J} \\
16 & J1844$-$0128    &  30.81  & 0.90 & 0.2349& \cite{2025RAA....25a4003W} \\
17 & J1018$-$1523$^{\ast}$    & 257.13  & 33.53 &0.2277 & \cite{Swiggum+2023ApJ...944..154S} \\
18 & J1756$-$2251    & 6.50   & 0.95  & 0.1806 & \cite{Ferdman+2014MNRAS.443.2183F} \\
19 & J1411+2551      & 33.38    & 72.10  & 0.1699& \cite{Martinez+2017ApJ...851L..29M} \\
20 & J1829+2456      & 53.34   & 15.61  & 0.1391& \cite{Haniewicz+2021MNRAS.500.4620H} \\
21 & J0453+1559      &  184.13    &  $-$17.14 & 0.1125& \cite{Martinez+2015ApJ...812..143M} \\
22 & J1755$-$2550$^{\ast}$    &  3.80   &-0.32 & 0.08935 & \cite{2018MNRAS.476.4315N} \\
23 & J0737$-$3039B   &  245.24&   $-$4.50  & 0.08778 & \cite{Kramer+2021PhRvX..11d1050K} \\
23 & J0737$-$3039A   & 245.24   & $-$4.50 & 0.08778 & \cite{Kramer+2021PhRvX..11d1050K} \\
24 & J1325$-$6253    & 306.75 &  $-$0.27   & 0.06401 & \cite{Sengar+2022MNRAS.512.5782S} \\
\bottomrule
\end{tabular}
\end{table*}

\bsp
\label{lastpage}
\end{document}